\documentclass[reprint,aps,prl,longbibliography,twocolumn,floatfix,10pt]{revtex4-1}

\usepackage{bm}
\usepackage{dcolumn}% Align table columns on decimal point
\usepackage{amsthm}
\usepackage{amssymb}
\usepackage{amsmath}
\usepackage{graphicx}
\usepackage{bm}
\usepackage{xcolor}
\usepackage{fix-cm}
\usepackage{mathptmx} % rm & math
\usepackage[T1]{fontenc}
\usepackage[colorlinks,allcolors=blue]{hyperref}
\makeatletter
\def\NAT@def@citea{\def\@citea{\NAT@separator}}
\makeatother

\begin{document}

\title{Timescales of quantum equilibration, dissipation and fluctuation in nuclear collisions}

% Force line breaks with \\
\author{C. Simenel}\email{cedric.simenel@anu.edu.au}
\affiliation{Department of Theoretical Physics and Department of Nuclear Physics, Research School of Physics and Engineering, The Australian National University, Canberra ACT  2601, Australia}
\author{K. Godbey}\email{kyle.s.godbey@vanderbilt.edu}
\author{A.S. Umar}\email{umar@compsci.cas.vanderbilt.edu}
\affiliation{Department of Physics and Astronomy, Vanderbilt University, Nashville, Tennessee 37235, USA}
\date{\today}

%------------------------------------------------------------------------------

\begin{abstract}
% the extra lines centers the abstract like the final PRC document
\edef\oldrightskip{\the\rightskip}
%\begin{description}
\rightskip\oldrightskip\relax
\setlength{\parskip}{0pt} % no skip between items
%\item[Background] 
Understanding the dynamics of equilibration processes in quantum systems as well as their interplay with dissipation and fluctuation is a major challenge in quantum many-body theory. 
%\item[Purpose] 
The timescales of such processes are investigated in collisions of atomic nuclei using fully microscopic approaches. 
%\item[Methods] 
Results from time-dependent Hartree-Fock (TDHF) and time-dependent random-phase approximation (TDRPA) calculations are compared for 13 systems over a broad range of energies. %, with collision partners from $^{40}$Ca to $^{249}$Bk, and at various energies from about 10\% below to more than twice the Coulomb barrier. 
%\item[Results] 
The timescale for full mass equilibration ($\sim2\times10^{-20}$s) is found to be much larger than timescales for neutron-to-proton equilibration, kinetic energy and angular momentum dissipations which are on the order of $10^{-21}$s. Fluctuations of mass numbers in the fragments and correlations between their neutron and proton numbers build up within only a few $10^{-21}$s. 
%\item[Conclusions] 
This indicates that dissipation is basically not impacted by mass equilibration, but is mostly driven by the exchange of nucleons between the fragments. 
%\end{description}
\end{abstract}
\maketitle

%------------------------------------------------------------------------------

%\section{Introduction}

It is well known that two classical systems with differing properties (e.g., two fluids of different colors or temperatures) that are able to exchange matter in some way tend to equilibrate their initial asymmetry over time. 
This equilibration process also occurs in quantum systems, a phenomenon actively studied at present with 
quantum simulations using ultracold atoms and trapped ions~\cite{eisert2015}.
The principal challenge is in characterizing the path followed by a quantum system as it evolves towards equilibrium. 
Along this path, the system also encounters dissipation of collective energy (e.g., kinetic energy of collision partners)
into internal degrees of freedom. 
Although equilibration processes are expected to depend on dissipative mechanisms, 
the interplay between equilibration and dissipation in quantum systems is still not well understood.

Collisions of atomic nuclei are ideal to investigate equilibration and dissipative processes in quantum many-body systems~\cite{moretto1981,viola1987}. 
Indeed, these collisions are usually too fast (on the order of a few zeptoseconds, 1~zs$=10^{-21}$~s) to allow interaction with external environment to affect the outcome of the collision.
For example, emission of $\gamma$ photons due to coupling to the electromagnetic field occurs over a much longer timescale, usually greater than $10^{-18}$~s. 
Nevertheless, nucleons, which move at about 20\% speed of light in the nucleus can be transferred from one nucleus to the other when the nuclei are in contact. 
As a result, the internal degrees of freedom characterising the nuclear states encounter a rapid rearrangement, tending to equalize initial asymmetries between the collision partners as well as dissipating their kinetic energy and angular momentum. 
In addition, the relatively small number of nucleons at play (up to a few hundred) makes the problem numerically tractable if one makes relevant  approximations to the quantum many-body problem.

%Discussion of various asymmetries (mass, isospin) and dissipations (kinetic, angular momentum) at play. 

The mass (number of nucleons) of the nuclei and the difference between their neutron and proton numbers are among the main collective degrees of freedom which can encounter equilibration in heavy-ion collisions. 
A broad range of masses, and thus chemical potentials, are already accessible with stable beams. 
Moreover, the recent development of exotic beams has significantly increased the range of available asymmetries between protons and neutrons, leading to ambitious reaction mechanism programs at exotic beam facilities around the world, including FRIB (US)~\cite{glasmacher2017}, RIKEN-RIBF (Japan)~\cite{sakurai2010}, SPIRAL2 (France)~\cite{lewitowicz2011}, and FAIR-NUSTAR (Germany)~\cite{kalantar2018}.
In particular, neutron-proton asymmetric collisions are expected to bring valuable information on the density dependence of the nuclear symmetry energy, which is highly relevant in nuclear astrophysics (see~\cite{mcintosh2019} for a recent review).

As mass and neutron-proton equilibration occurs via a flow of nucleons between colliding partners, they are expected to be significantly impacted by nuclear viscosity. The latter is also responsible for dissipation of both the initial kinetic energy and the angular momentum of the fragments~\cite{randrup1982}. In turn, as a manifestation of Einstein's fluctuation-dissipation theorem, the multinucleon transfer between the fragments in contact is expected to build up quantum fluctuations that lead to broad distributions of particle numbers in the final fragments~\cite{randrup1978}.
Equilibration, dissipation, and fluctuation then form a complex network of interrelated observables. 

Investigating the interplay between these quantities requires advanced theoretical descriptions. 
We adopt fully microscopic time-dependent approaches to the nuclear many-body problem allowing for parameter-free (except for the underlying nuclear interaction) dynamical descriptions of the relevant quantities (see~\cite{simenel2012,simenel2018,sekizawa2019} for recent reviews). 
The number of transferred nucleons, the kinetic energy of the fragments, and their angular momenta are all described by one-body observables. Average values of these quantities are computed with the time-dependent Hartree-Fock (TDHF) mean-field theory which is optimized for expectation values of one-body operators~\cite{balian1981}. 
In particular, TDHF contains all one-body dissipation mechanisms which are the most relevant at the energies considered in this work (see, e.g., discussion in~\cite{simenel2012}).
However, TDHF often underestimates fluctuations of these operators~\cite{koonin1977,dasso1979}. 
We thus compute fluctuations of neutron and proton numbers (as well as their correlations) in the fragments from the time-dependent random-phase approximation (TDRPA) prescription of Balian and V\'en\'eroni~\cite{balian1984} which is indeed optimized on fluctuations of one-body operators in the limit of small fluctuations. 

Probably the most important quantity characterizing various equilibration, dissipation, and fluctuation processes is the timescale over which they occur. 
A first step is then to investigate these timescales, their potential dependence on the entrance channel, and compare them to understand their relationships. 
Indeed, mechanisms with very different timescales are unlikely to be correlated while similar timescales indicate a potential common origin in the underlying microscopic mechanisms.
In this letter we present a systematic theoretical study of timescales for mass and neutron-proton equilibration and fluctuations, as well as kinetic energy and angular momentum dissipation.
TDHF and TDRPA results are presented for many collisions spanning a broad range of masses, energies and angular momenta (see supplemental material \cite{supplemental}).
Although most results are compiled from published data, we have performed new calculations for completeness with the TDHF3D code~\cite{kim1997} for $^{40}$Ca$+^{40,48}$Ca,$^{64}$Ni (see, e.g.,~\cite{simenel2008} for numerical details) and with the code of Ref.~\cite{umar2006c} for $^{176}$Yb$+^{176}$Yb. These new calculations used the SLy4$d$ parametrization of the energy density functional~\cite{kim1997}.

%\section{Method}\label{sec:method}

In order to compare various systems with different initial conditions, let us introduce a generic way of defining a ``normalized'' observable $\delta X(\tau)=\frac{X(\tau)-X_\infty}{X_0-X_\infty}$ where $X(\tau)$ is the quantity used to characterise equilibration, dissipation or fluctuations. 
It is a function of the contact time $\tau$ between the fragments before they re-separate. 
Here, contact is usually defined by two fragments linked by a neck, with a neck density exceeding half the saturation density $\rho_{sat}/2\simeq0.08$~fm$^{-3}$.
The initial value of $X$ is noted $X_0$.
For long contact times, the value of $X$ is expected to saturate to its equilibrium value $X_\infty$, in which case $\delta X \rightarrow 0$. 
For quasielastic collisions in which contact does not occur (with the above definition of contact), the contact time is obviously
$\tau=0$, leading to $\delta X(0)=1$. 
However, the nuclei may interact before contact through the overlap of their density tails, 
possibly leading to values of $\delta X(0)\ne1$.
%\section{Results}\label{sec:results}

\begin{figure}[!htb]
\centerline{\includegraphics*[width=8.6cm]{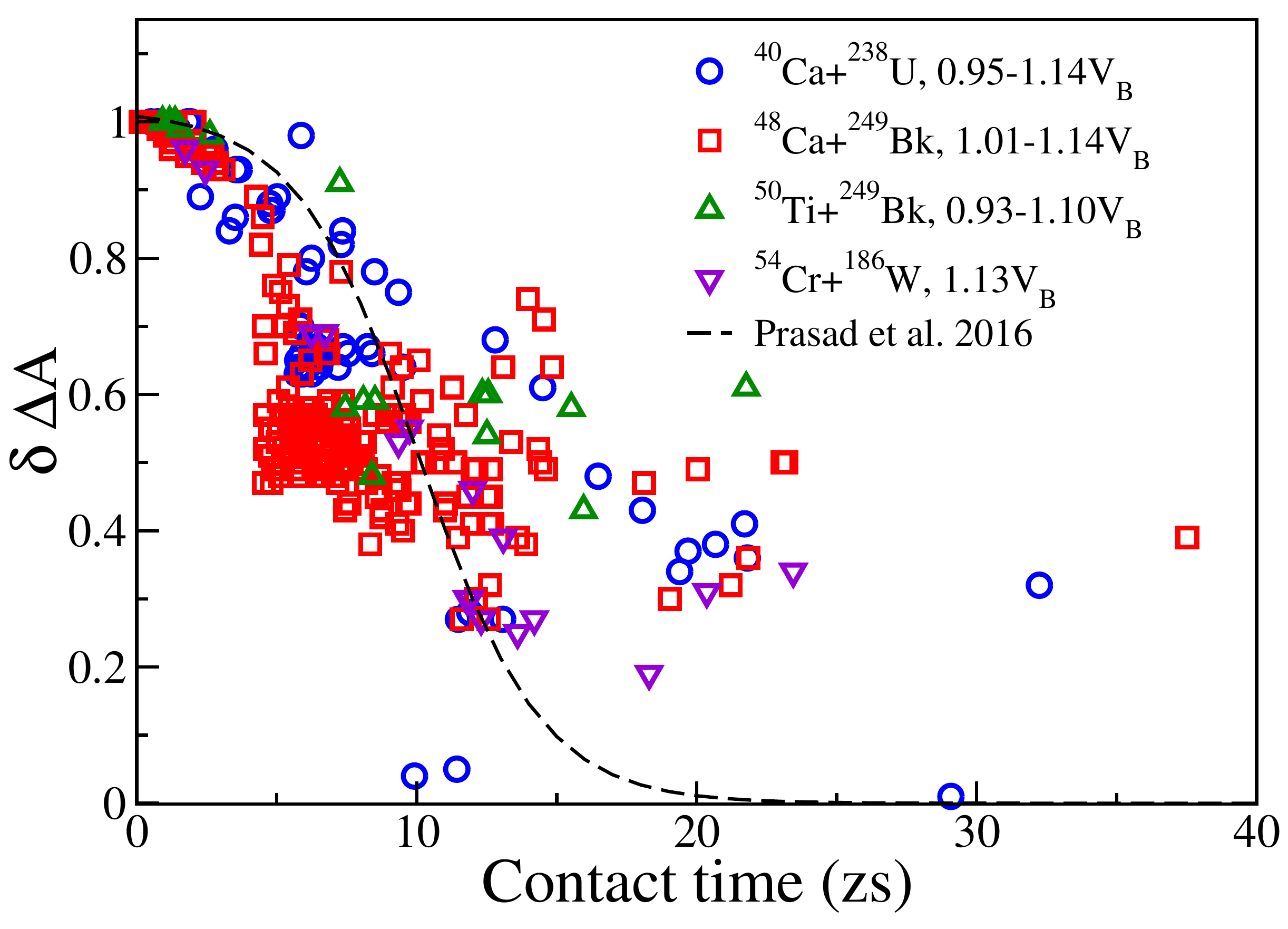}}
\caption{\protect(Color online) Fragment mass asymmetry as a function of contact times from TDHF calculations of $^{40}$Ca$+^{238}$U~\cite{oberacker2014}, $^{48}$Ca,$^{50}$Ti$+^{249}$Bk~\cite{umar2016,godbey2019}, and $^{54}$Cr$+^{186}$W~\cite{hammerton2015}. Energy ranges are given as function of the barrier height $V_B$~\cite{swiatecki2005}. The dashed line shows the expected equilibration assuming Fermi-type mass drift determined experimentally by Prasad \textit{et al.}~\cite{prasad2016}.}
\label{fig:A}
\end{figure}

Equilibration of mass asymmetry $\delta\,\Delta A(\tau)$ is studied in Fig.~\ref{fig:A} with $\Delta A\equiv A_1-A_2$ and $A_{1,2}$ the number of nucleons in the outgoing fragments obtained from a series of TDHF calculations~\cite{oberacker2014,umar2016,godbey2019,hammerton2015} for systems at energies near the Coulomb capture barrier $V_B$~\cite{swiatecki2005}. 
For each system and energy, a range of angular momenta is considered, thus producing a distribution of contact times with various exit channels. 
Note that each point typically requires several days of computational time on modern computers. 
The equilibrium value for mass asymmetry is chosen to be $\Delta A_\infty=0$, i.e., with two outgoing fragments of similar masses. 
Note that this equilibrium is rarely reached due to shell effects in the fragments favoring exit channels before full symmetry is achieved~\cite{wakhle2014,sekizawa2016,godbey2019}.
Despite large fluctuations of $\delta\,\Delta A(\tau)$, all systems exhibit a similar pattern, with an equilibration of mass asymmetry and a full symmetry expected to be reached at about 20~zs contact time in average. 
This indicates that mass equilibration is a relatively slow process.
Such reactions are called quasifission~\cite{toke1985} as fission-like fragments are produced without formation of an intermediate compound nucleus (which would require much longer time). These TDHF predictions of mass equilibration times are in good agreement with ``neutron clock''~\cite{hinde1992} and fragment mass-angle distribution~\cite{toke1985,durietz2013,prasad2016} measurements. 

\begin{figure}[!htb]
\centerline{\includegraphics*[width=8.6cm]{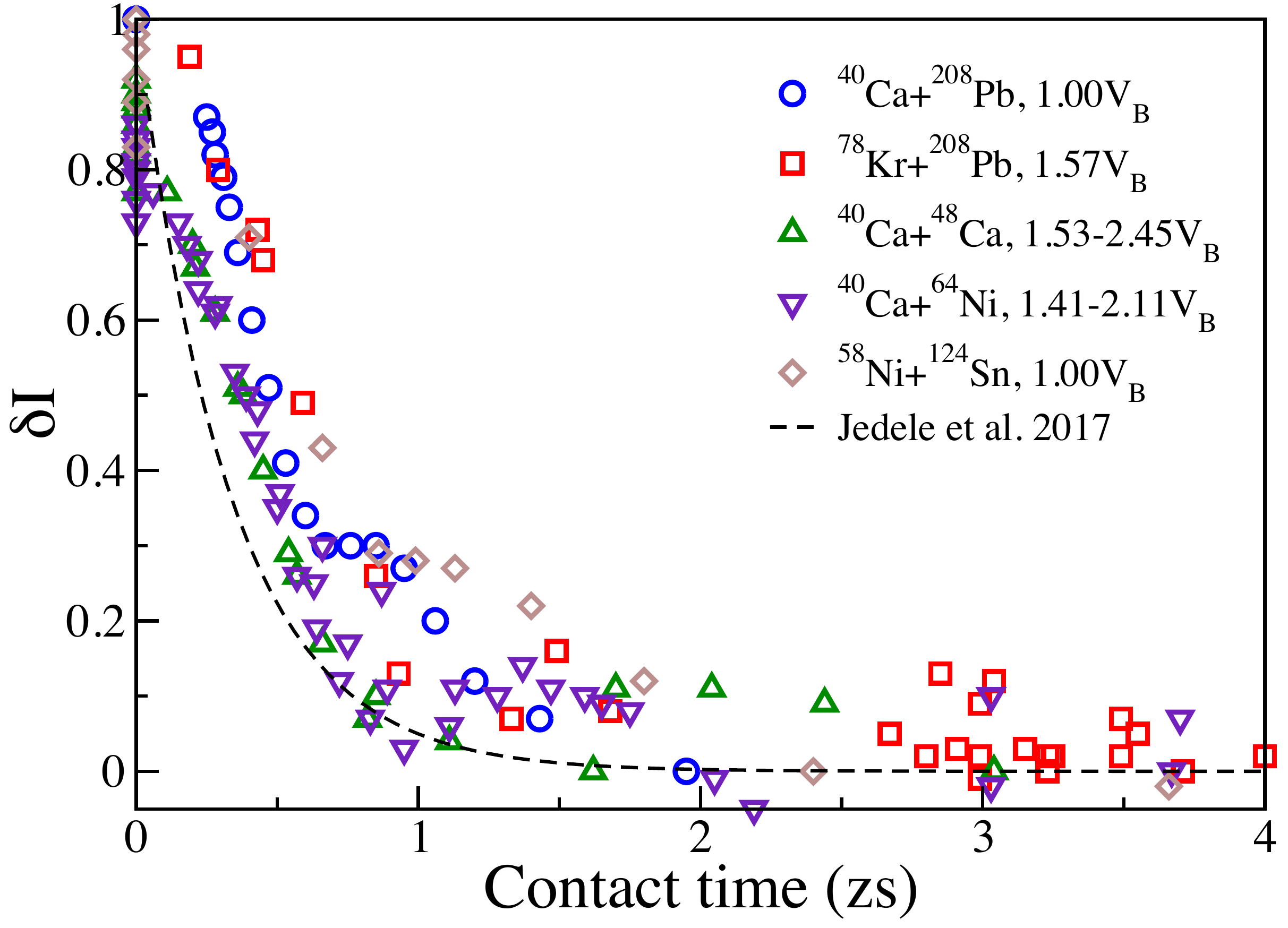}}
\caption{\protect(Color online) Asymmetry between proton and neutron numbers as a function of contact times from TDHF calculations of $^{40}$Ca$+^{208}$Pb~\cite{simenel2012b}, $^{78}$Kr$+^{208}$Pb~\cite{umar2017}, $^{40}$Ca$+^{48}$Ca,$^{64}$Ni (this work),  and $^{58}$Ni$+^{124}$Sn~\cite{wu2019}. Energy ranges are given as function of the barrier height $V_B$~\cite{swiatecki2005}.
The dashed line shows the expected equilibration assuming the rate constant of 3~zs$^{-1}$ determined experimentally by Jedele \textit{et al.}~\cite{jedele2017}.}
\label{fig:I}
\end{figure}

We now investigate equilibration of initial asymmetry between proton and neutron numbers, quantified by $I=(N_1-Z_1)-(N_2-Z_2)$, with $Z_{1,2}$ the number of protons and $N_{1,2}=A_{1,2}-Z_{1,2}$ the number of neutrons in the fragments. 
%By convention, we choose to assign the subscript 1 to the nucleus with the largest initial value of $N-Z$. 
Figure~\ref{fig:I} shows $\delta I(\tau)$ for various systems studied with TDHF~\cite{simenel2012b,umar2017,wu2019}.
Here, the equilibrium values $I_\infty$ are smaller (in magnitude) than $I_0$, though not necessarily zero, and $I_\infty$ needs to be determined for each system. 
This is done by taking $I(\tau)$ for large times when it does not significantly change with $\tau$ anymore. 
The fact that we usually have $I_\infty\ne0$ (see, e.g., Fig.~10 of Ref.~\cite{umar2017}) is essentially due to the remaining charge ($Z$) asymmetry  which favors different values of $N-Z$ in the fragments because of different Coulomb energies.
We see in Fig.~\ref{fig:I} that neutron-proton equilibration is a much faster phenomenon than mass equilibration as $\delta I(\tau)$ is approximately zero for $\tau\sim1$~zs and above. 
This timescale is in good agreement with what has been recently determined experimentally from reactions at intermediate energies~\cite{jedele2017} (dashed line in Fig.~\ref{fig:I}). 

\begin{figure}[!htb]
\centerline{\includegraphics*[width=8.6cm]{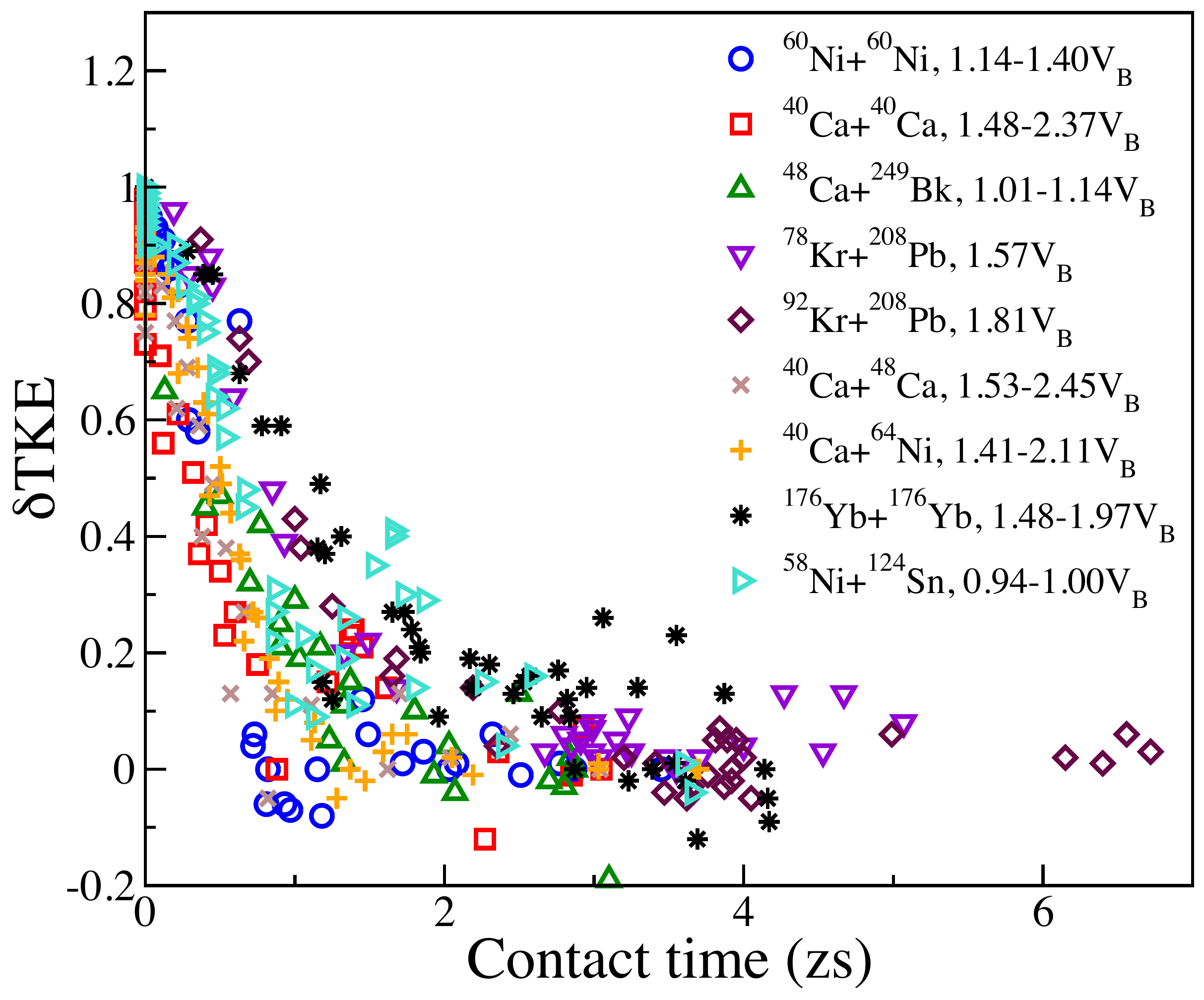}}
\caption{\protect(Color online) Evolution of $\delta$TKE as a function of contact times from TDHF calculations of $^{60}$Ni$+^{60}$Ni~\cite{williams2018}, $^{40}$Ca$+^{40}$Ca~\cite{simenel2011} (and this work), $^{48}$Ca$+^{249}$Bk~\cite{umar2016,godbey2019}, $^{78,92}$Kr$+^{208}$Pb~\cite{umar2017}, $^{40}$Ca$+^{48}$Ca,$^{64}$Ni (this work), $^{176}$Yb+$^{176}$Yb~\cite{godbey2020b} (and this work), and $^{58}$Ni$+^{124}$Sn~\cite{wu2019}. Energy ranges are given as function of the barrier height $V_B$~\cite{swiatecki2005}.}
\label{fig:E}
\end{figure}

Next, we investigate the timescale for dissipation of total kinetic energy (TKE) of the fragments. 
This usually involves reactions well above the barrier, in which collision terms (not included in 
TDHF) could affect the reaction mechanisms. Nevertheless, the fact that fully damped collisions
are obtained in TDHF \cite{williams2018,simenel2011,umar2016,umar2017,godbey2019,wu2019,sekizawa2019}, together with comparison between TKE predictions and experimental data \cite{williams2018},
 indicate that one-body dissipation mechanisms (the only ones included in TDHF) are sufficient up to these energies. 
Fully damped collisions have a value of TKE$_\infty$ which depends on the system.
It roughly corresponds to the Coulomb repulsion between the fragments at scission, and is usually well approximated by Viola systematics~\cite{viola1985,hinde1987}.
However, the final TKE for a given system can also be affected by initial conditions such as the orientation of a deformed collision partner in the entrance channel~\cite{umar2016}.
Therefore, we determine the value of TKE$_\infty$ for each system and each energy (when a broad range
 of energies is considered). 
%on a case-by-case basis, as in the case of neutron-to-proton equilibration. 

The resulting  $\delta$TKE$(\tau)$, as predicted by TDHF calculations~\cite{williams2018,simenel2011,umar2016,umar2017,godbey2019,wu2019}, are shown in Fig.~\ref{fig:E}. 
A full dissipation is obtained after typically $1-2$~zs, indicating a fast process with similar timescale as in neutron-to-proton equilibration. 
Experimentally, a possible way to extract timescales is through fragment angular distributions. 
Recent TDHF predictions of TKE-angle correlations in $^{58}$Ni+$^{60}$Ni deep-inelastic collisions have been found to be in good agreement with experiment~\cite{williams2018}, supporting our extracted timescale for energy dissipation. 

\begin{figure}[!htb]
\centerline{\includegraphics*[width=8.6cm]{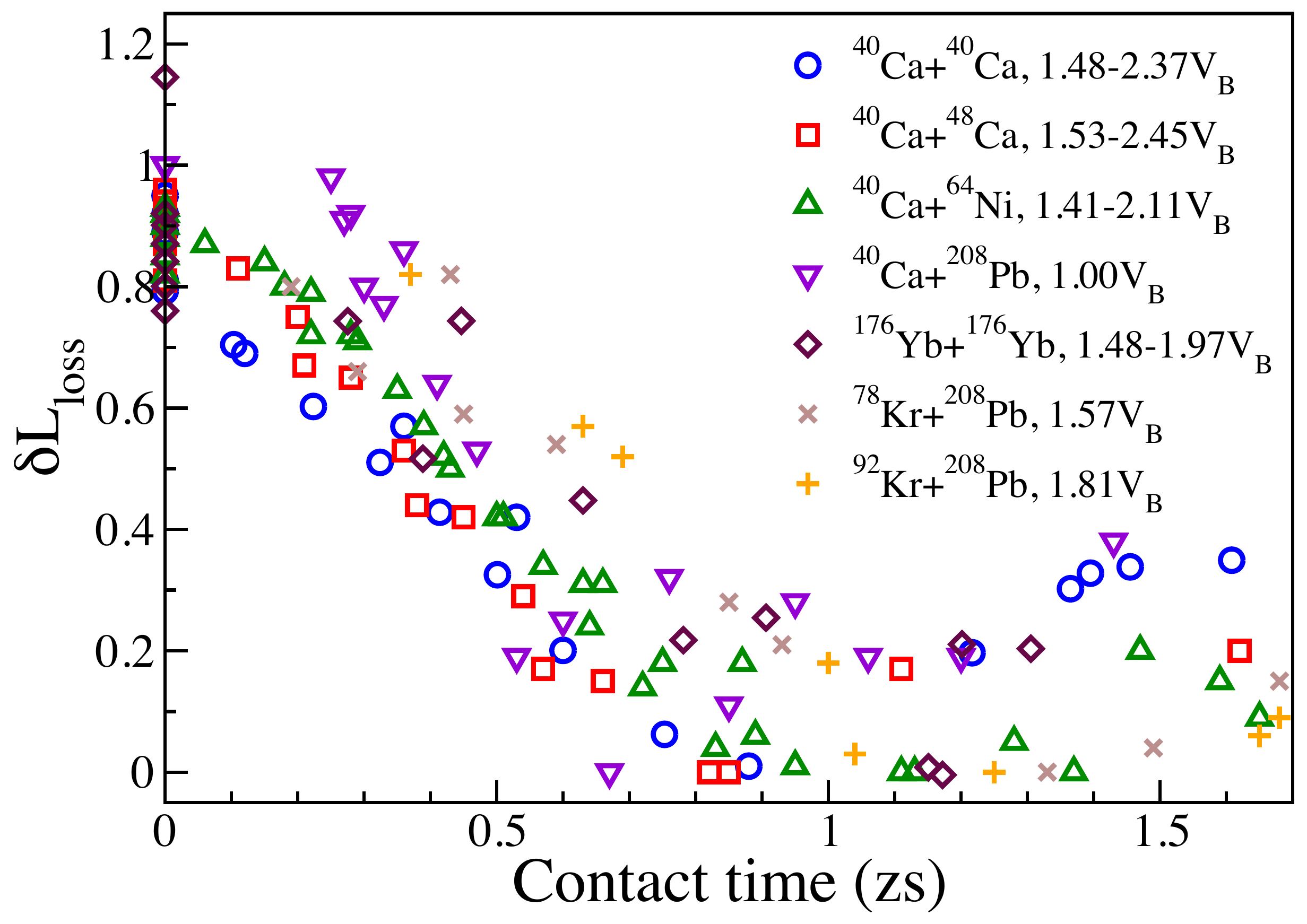}}
\caption{\protect(Color online) Evolution of $\delta L_{loss}$ as a function of contact times from TDHF calculations of $^{40}$Ca$+^{40}$Ca~\cite{simenel2011} (and this work), $^{40}$Ca$+^{48}$Ca,$^{64}$Ni (this work), $^{40}$Ca$+^{208}$Pb~\cite{simenel2012}, $^{176}$Yb+$^{176}$Yb~\cite{godbey2020b} (and this work), and $^{78,92}$Kr$+^{208}$Pb~\cite{umar2017}. Energy ranges are given as function of the barrier height  $V_B$~\cite{swiatecki2005}.}
\label{fig:L}
\end{figure}

We also investigated dissipation of orbital angular momentum $L$ between the fragments. 
Let us define $L_{loss}(\tau)=L_0-L(\tau)$ as the difference between initial and final angular momentum, giving ${L_{loss\,_0}}=0$. 
The behaviour of $L_{loss}(\tau)$ for a given system at a given energy (i.e., varying only $L_0$), is first to rise with $\tau$ due to dissipation, 
and then to decrease slowly at large contact times, as the latter correspond to more central collisions, i.e. with less initial angular momentum  $L_0$ available for dissipation. 
We thus define the equilibrium value ${L_{loss_\infty}}$ as the maximum value of $L_{loss}(\tau)$. 
Figure~\ref{fig:L} shows the resulting evolution of $\delta L_{loss}$ with contact time for a set of reactions studied with TDHF~\cite{simenel2011,simenel2012,godbey2020b,umar2017}.
Note that $\delta L_{loss}>1$ is sometimes observed for collisions of deformed nuclei that allow transfer from intrinsic to orbital angular momentum.
Nevertheless, most reactions have dissipated their angular momentum within 1~zs.

\begin{figure}[!htb]
\centerline{\includegraphics*[width=8.6cm]{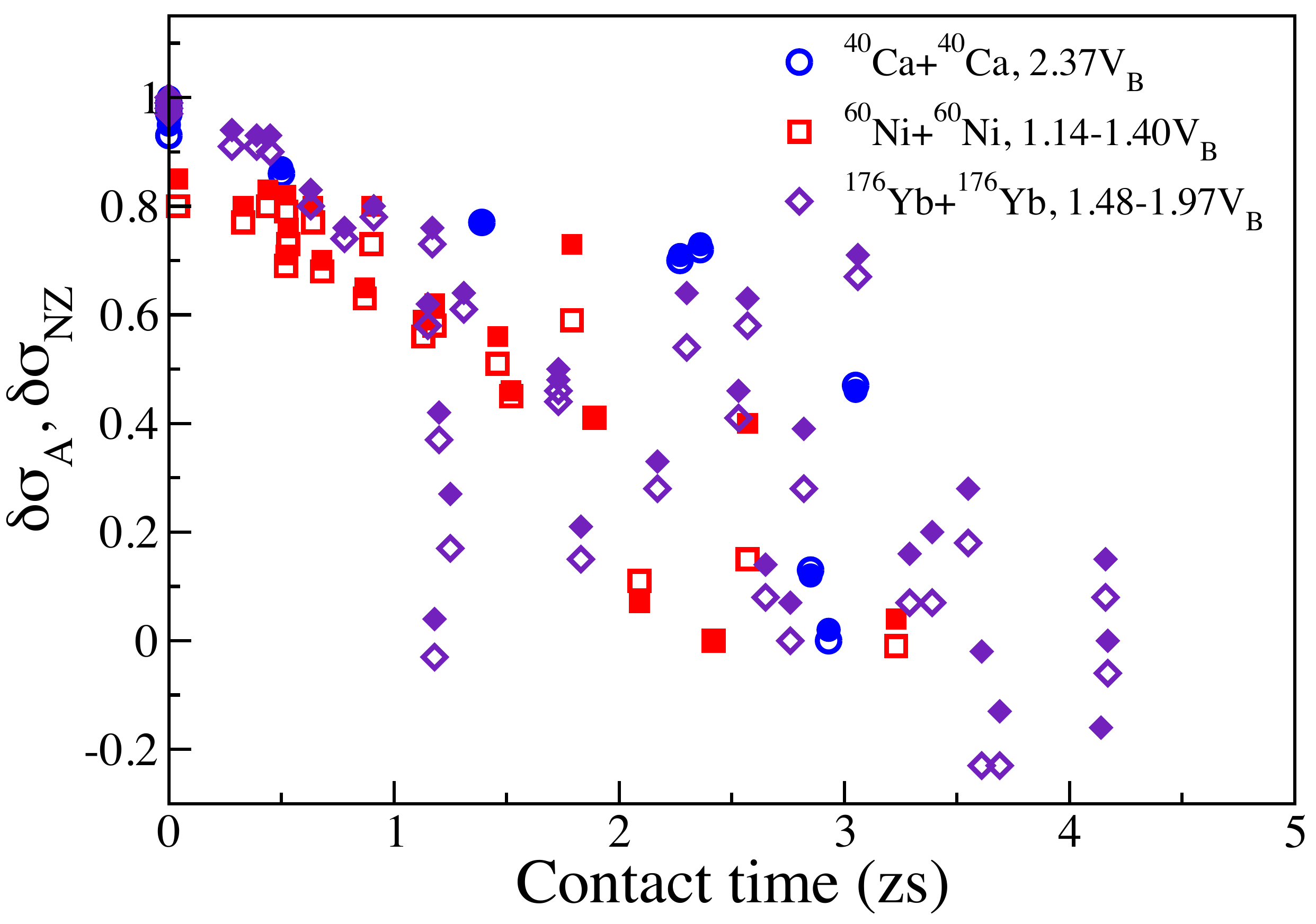}}
\caption{\protect(Color online) Evolution of $\delta \sigma_A$ associated with mass fluctuations (open symbols) and $\delta \sigma_{NZ}$ associated with neutron-proton correlations (full symbols) as a function of contact times from TDRPA calculations of $^{40}$Ca$+^{40}$Ca~\cite{simenel2011} (and this work), $^{60}$Ni$+^{60}$Ni~\cite{williams2018}, and $^{176}$Yb+$^{176}$Yb~\cite{godbey2020b} (and this work). Energy ranges are given as function of the barrier height $V_B$~\cite{swiatecki2005}.}
\label{fig:S}
\end{figure}

Let us finally study the dynamics of mass fluctuations $\sigma_A=\sqrt{\langle\hat{A}^2\rangle-\langle\hat{A}\rangle^2}$ and proton-neutron correlations $\sigma_{NZ}=\sqrt{\langle\hat{N}\hat{Z}\rangle-\langle\hat{N}\rangle\langle\hat{Z}\rangle}$, where $\hat{A}$, $\hat{N}$ and $\hat{Z}$ count particles in the fragments. 
These fluctuations and correlations have been determined with TDRPA in deep inelastic collisions (DIC)~\cite{simenel2011,williams2018,godbey2020b}. These systems are symmetric to avoid possible spurious effects due to skewed fragment mass distributions, which may occur in asymmetric systems, and which prevent the interpretation of TDRPA results as fluctuations~\cite{williams2018}. 
As the particle numbers in the incoming fragments are well defined, we have $\sigma_{A_0}=\sigma_{NZ_0}=0$. 
It is also found that, in most central DIC, TDRPA fluctuations exhibit relatively large fluctuations around an average value which we choose for $\sigma_{A,NZ\,_\infty}$. 

The evolutions of $\delta\sigma_A(\tau)$ and $\delta\sigma_{NZ}(\tau)$ as a function of contact time are shown in Fig.~\ref{fig:S}.
The observed decrease of $\delta\sigma_A(\tau)$ and $\delta\sigma_{NZ}(\tau)$ indicates that fluctuations and correlations build up within a timescale of about 3~zs. 
Naturally, the choices of $\sigma_{A_\infty}$ and $\sigma_{{NZ}_\infty}$ are quite arbitrary, and changing these values would impact the slopes of $\delta\sigma(\tau)$. 
Nevertheless, as all three systems reach their maximum fluctuations at similar contact times, this would not have a significant influence on the resulting timescales. 
However, the small number of systems, together with the large variations of fluctuations due to the sensitivity to initial condition in DIC, prevent to draw definitive conclusions with respect to the universality of the behaviour of mass fluctuations. More studies would be welcome.

%\section{Discussion}\label{sec:discussion}

A comparison between the different timescales for equilibration, dissipation and fluctuation processes leads to several conclusions. 
First, it is quite remarkable that each process exhibits similar timescales despite being derived from extremely different systems and energies. 
This is a strong indication that the underlying mechanisms are universal and do not depend much on the specifics of the entrance channel. 
Note that the present systematics also include calculations by several teams with different TDHF \cite{kim1997,umar2006c,maruhn2014,schuetrumpf2018,sekizawa2013} and TDRPA \cite{simenel2011,williams2018,godbey2020b} codes and two different energy density functionals \cite{kim1997,chabanat1998a}, showing the robustness of the results.

Second, it is noticeable that neutron-to-proton equilibration occurs on very similar timescales as angular momentum and kinetic energy dissipations. 
This points toward a strong correlation between these mechanisms. 
However, neutron-to-proton equilibration cannot be the sole dissipative mechanism as collisions in systems with neutron-to-proton symmetry have a similar timescale for dissipation (see Fig.~\ref{fig:E}). 

The main mechanism for dissipation is expected to be multinucleon transfer between the fragments~\cite{randrup1978}. 
The least bound nucleons (closest to the Fermi surface) are usually more freely transferred. 
In neutron-to-proton asymmetric systems, this essentially leads to protons flowing one way and neutrons being transferred the other way. 
In the case of symmetric systems, however, protons and/or neutrons flow both ways. If the corresponding timescale is the same as for neutron-to-proton equilibration, this would explain why dissipation timescales are the same in neutron-to-proton symmetric and asymmetric systems. 

Another observation is that mass equilibration is much slower (by more than one order of magnitude) than dissipation.
It takes place in systems in which energy and angular momentum have already been damped. 
From this separation of time scales we can conclude that mass equilibration is not expected to be a significant contributor to dissipation processes. 

Finally, the fluctuations and correlations in particle numbers build up within a few zeptoseconds, which is a bit slower than dissipation. Nevertheless, it is still much faster than mass equilibration. 
Note, however, that TDRPA predictions of fluctuations are only available for symmetric systems. 
The increase of mass fluctuations with contact time in symmetric collisions is a clear signature that multinucleon transfer has happened (both ways due to symmetry) within this time frame. 
It would be interesting to investigate timescales for fluctuations and correlations
 in asymmetric systems with beyond TDRPA models such as the stochastic mean-field approach~\cite{lacroix2014}.

%------------------------------------------------------------------------------
%\clearpage
\begin{acknowledgments}
We thank D. J. Hinde for useful discussions.
This work has been supported by the Australian Research Council Discovery Project (project numbers DP160101254 and DP190100256) funding schemes and by the
 U.S. Department of Energy under grant No. DE-SC0013847 with Vanderbilt University.
The calculations have been performed in part at the NCI National Facility in Canberra, Australia, 
which is supported by the Australian Commonwealth Government.
\end{acknowledgments}

%------------------------------------------------------------------------------
\bibliography{VU_bibtex_master.bib}

\onecolumngrid
\appendix

\newpage
\section*{Supplemental online material for:
Dynamics of quantum equilibration, dissipation and fluctuation in nuclear collisions}

%\begin{abstract}
This supplemental material provides details of the time-dependent Hartree-Fock and time-dependent RPA results used to produce the figures of the main manuscript.
%\end{abstract}

\maketitle

\subsection{Generalities}
The results presented in this supplemental material are for TDHF calculations, except for fluctuations $\sigma_A$ and correlations $\sigma_{NZ}$ which have been obtained from TDRPA calculations. 
The data are for primary fragments, i.e., prior to subsequent decay.
Energies are in MeV,  angular momenta in units of $\hbar$, and contact times in zeptoseconds (1~zs$=10^{-21}$~s).
The TKE values are for final total kinetic energy of the fragments. 
$L_{loss}$ refers to the loss of total orbital angular momentum between the fragments. 

Data are only presented for calculations leading to two fragments in the exit channel. 
It is usually assumed that fusion happens when only one fragment is present at the end of the TDHF calculation, which occurs for initial orbital angular momentum smaller than the lowest value of $L_0$ reported when the latter is non-zero. 

Each section presents the results for a specific system, with a brief description of the calculations at the beginning. 
The subscript $H$ and $L$ stand for heavy and light fragment, respectively. 
The extracted values of $X_\infty$  are also given (see definition in main manuscript).

\newpage
\subsection{$^{40}\mbox{Ca}+^{40}\mbox{Ca}$}

TDHF and TDRPA calculations at $E_{c.m.}=128$~MeV are from \cite{simenel2011} with the TDHF3D code and the SLy4$d$ parametrization of the Skyrme interaction \cite{kim1997}.
TDHF calculations at $E_{c.m.}=80$ and 100~MeV have been produced for the present work with the same code and functional.  
The Coulomb capture barrier for this system is estimated to be $V_B=54.1$~MeV \cite{swiatecki2005}.

 \begin{table}[ht]
% \caption{\label{tab:tip} }
 \begin{ruledtabular}
 \begin{tabular}{lcccccc}
$E_{c.m.}$ & $L_0$ & $\tau$ & TKE & $L_{loss}$ & $\sigma_A$ & $\sigma_{NZ}$\\
\colrule
%E    Li     T	    TKE     Lf    SA	 SNZ
 80&     50&   1.04&        64.12&        6.92\\     
&          51&  0.27&         65.79&        6.04\\     
&          52&   0&        72.96&        2.84\\
&          53&   0&        75.79&        1.50\\  
&          54&   0&        77.05&       0.96\\    
&          55&   0&        77.78&       0.55    \\
 \colrule
100& 64&   0.88&  64.24&  19.07& &\\
&65&   0.53&  72.58 & 11.14&&\\
&66&   0.36&  77.50 &  8.35&&\\
&67&   0.12&  84.35 &  5.98&&\\
&68&   0&  90.26  & 3.60&&\\
&69&   0&  93.63 &  2.16&&\\
&70&   0&  95.46 &  1.51&&\\
\colrule
128& 67&  2.36&     68.97&      15.85&  4.544&   2.177\\
&     68&     2.27&     60.34&  15.35&  4.817&   2.293\\
&     69&     2.85&     66.79&  18.52& 14.124&   7.001\\
&     70&     3.05&     67.42&  20.44&  8.662&   4.283\\
&     72&     2.93&     71.67&  14.97& 16.313&   7.783\\
&     73&     1.61&     75.75&  12.34& -- & --\\
&     74&     1.45&     79.96&  13.90&--&--\\
&     75&     1.39&     81.71&  13.53&  3.768&   1.785\\
&     76&     1.37&     81.16&  13.89&--&--\\
&     77&     1.22&     76.48&  15.76&--&--\\
&     78&     0.75&     78.39&  18.32&--&--\\
&     79&     0.60&     83.67&  16.08&--&--\\
 &    80&     0.50&     87.98&  13.64&  2.276&   1.016\\
 &    81&     0.41&     92.77&  11.49&--&--\\
 &    82&     0.32&     98.08&   9.90&--&--\\
 &    83&     0.22&    104.21&   8.19&--&--\\
 &    84&     0.10&    110.20&   6.18&--&--\\
 &    85&     0&    114.98&      4.48&  1.118&   0.390\\
 &    90&     0&    124.25&      1.37&  0.538&   0.106\\
 &    95&     0&    126.02&      0.79&  0.320&   0.039\\
 &    100&    0&    126.56&      0.62&  0.204&   0.017
  \end{tabular}
 \end{ruledtabular}
 \end{table}

 \begin{table}[ht]
 %\caption{\label{tab:tip}}
 \begin{ruledtabular}
 \begin{tabular}{lcccc}
$E_{c.m.}$ & TKE$_\infty$ &  $L_{loss_\infty}$ & $\sigma_{A_\infty}$ & $\sigma_{NZ_\infty}$\\
\colrule
80 &  64.1&  6.9  & -- &--\\
100 &  64.2&  19.1   & -- &--\\
128& 67.4 & 18.3 & 16.3 & 7.8    
  \end{tabular}
 \end{ruledtabular}
 \end{table}

\newpage
\subsection{$^{40}\mbox{Ca}+^{48}\mbox{Ca}$}

TDHF calculations for this system have been made for this work with the TDHF3D code and the SLy4$d$ functional \cite{kim1997}.
The Coulomb capture barrier for this system is estimated to be $V_B=52.3$~MeV \cite{swiatecki2005}.

 \begin{table}[ht]
% \caption{\label{tab:tip} }
 \begin{ruledtabular}
 \begin{tabular}{lcccccccc}
$E_{c.m.}$ & $L_0$ & $\tau$ & TKE & $L_{loss}$ & $Z_H$ & $N_H$ & $Z_L$ & $N_L$\\
\colrule
80&55&   0.71&  60.28&   8.74&  21.48&  26.81&  18.52&  21.19\\
&56&   0.27&  69.97&   4.17&  20.76&  27.47&  19.24&  20.53\\
&57&   0.00&  74.64&   1.64&  20.40&  27.71&  19.60&  20.29\\
&58&   0.00&  76.33&   0.83&  20.26&  27.80&  19.74&  20.20\\
&59&   0.00&  77.20&   0.47&  20.18&  27.85&  19.82&  20.15\\
&60&   0.00&  77.60&   0.35&  20.13&  27.89&  19.87&  20.11\\
\colrule
100&69&   1.62&  68.83&  10.48&  21.80&  26.48&  18.20&  21.52\\
&70&   0.82&  67.12&  13.05&  21.64&  26.56&  18.36&  21.44\\
&71&   0.57&  72.86&  10.83&  21.37&  26.93&  18.63&  21.07\\
&72&   0.38&  81.24&   7.25&  20.95&  27.27&  19.05&  20.73\\
&73&   0.21&  88.22&   4.28&  20.60&  27.50&  19.40&  20.50\\
&74&  0&  92.24&   2.51&  20.41&  27.64&  19.59&  20.36\\
&75&   0&  94.54&   1.52&  20.29&  27.74&  19.71&  20.26\\
&76&   0&  95.82&   1.02&  20.22&  27.80&  19.78&  20.20\\
%&77&   0&  96.61&   0.80&  20.17&  27.84&  19.83&  20.16\\
%&78&   0&  97.16&   0.56&  20.13&  27.87&  19.87&  20.13\\
\colrule
128&80&   3.04&  62.07&  18.59&  22.69&  27.00&  17.31&  21.00\\
&81&   2.44&  66.07&  16.17&  22.18&  26.75&  17.82&  21.25\\
&82&   2.04&  63.60&  14.54&  22.35&  27.05&  17.65&  20.95\\
&83&   1.70&  70.47&  13.79&  22.31&  26.99&  17.69&  21.01\\
&84&   1.11&  69.61&  16.35&  22.07&  26.41&  17.93&  21.59\\
&85&   0.85&  70.83&  19.79&  21.86&  26.41&  18.14&  21.59\\
&86&   0.66&  79.79&  16.84&  21.70&  26.53&  18.30&  21.47\\
&87&   0.54&  87.31&  13.98&  21.48&  26.74&  18.52&  21.26\\
&88&   0.45&  94.24&  11.50&  21.23&  26.95&  18.77&  21.05\\
&89&   0.36& 100.94&   9.23&  20.99&  27.12&  19.01&  20.88\\
&90&   0.28& 107.32&   6.97&  20.77&  27.29&  19.23&  20.71\\
&91&   0.20& 112.74&   4.97&  20.60&  27.45&  19.40&  20.55\\
&92&   0.11& 116.68&   3.42&  20.46&  27.57&  19.54&  20.43\\
&93&   0& 119.39&   2.51&  20.36&  27.67&  19.64&  20.33\\
&94&   0& 121.22&   1.92&  20.28&  27.74&  19.72&  20.26\\
&95&   0& 122.57&   1.43&  20.22&  27.79&  19.78&  20.21
  \end{tabular}
 \end{ruledtabular}
 \end{table}

 \begin{table}[ht]
 %\caption{\label{tab:tip}}
 \begin{ruledtabular}
 \begin{tabular}{lcccc}
$E_{c.m.}$ & TKE$_\infty$ &  $L_{loss_\infty}$ & $Z_{H_\infty}$ & $N_{H_\infty}$\\
\colrule
128& 62.1 & 19.8 & 22.7 & 27.0    \\
%100 &  68.8 &  13.1  & 21.8 & 26.5 \\
%80 & 60.3 & 8.8 & 21.5 & 26.8  
100 &  68.8 &  13.1  &  &  \\
80 & 60.3 & 8.8 &  &   
  \end{tabular}
 \end{ruledtabular}
 \end{table}

\newpage
\subsection{$^{40}\mbox{Ca}+^{64}\mbox{Ni}$}

TDHF calculations for this system have been made for this work with the TDHF3D code and the SLy4$d$ functional \cite{kim1997}.
The Coulomb capture  barrier for this system is estimated to be $V_B=71.1$~MeV \cite{swiatecki2005}.

 \begin{table}[ht]
% \caption{\label{tab:tip} }
 \begin{ruledtabular}
 \begin{tabular}{lcccccccc}
$E_{c.m.}$ & $L_0$ & $\tau$ & TKE & $L_{loss}$ & $Z_H$ & $N_H$ & $Z_L$ & $N_L$\\
\colrule
100&61&   1.37&  74.94&   9.76&  28.67&  34.15&  19.33&  21.85\\
&62&   0.87&  77.43&   7.99&  29.04&  34.87&  18.96&  21.13\\
&63&   0.66&  80.40&   6.71&  29.04&  35.06&  18.96&  20.94\\
&64&   0.43&  86.71&   4.89&  28.74&  35.25&  19.26&  20.75\\
&65&   0.22&  92.03&   2.77&  28.50&  35.46&  19.50&  20.54\\
&66&   0&  94.60&   1.78&  28.37&  35.59&  19.63&  20.41\\
&67&   0&  96.02&   1.15&  28.28&  35.68&  19.72&  20.32\\
%&68&   0&  96.82&   0.83&  28.22&  35.74&  19.78&  20.26\\
%&69&   0&  97.41&   0.64&  28.18&  35.78&  19.82&  20.22
\colrule
125&77&   3.70&  77.68&  13.31&  28.70&  33.95&  19.30&  22.05\\
&78&   3.03&  77.54&  15.40&  27.83&  33.06&  20.17&  22.94\\
&79&  fusion& --& -- & -- &-- & -- &-- \\
&80&   2.19&  77.11&  10.90&  28.81&  33.72&  19.19&  22.28\\
&81&   1.13&  81.34&  14.36&  29.07&  34.51&  18.93&  21.49\\
&82&   0.89&  85.00&  13.51&  29.23&  34.69&  18.77&  21.31\\
&83&   0.75&  90.14&  11.72&  29.23&  34.86&  18.77&  21.14\\
&84&   0.63&  95.14&   9.97&  29.14&  35.02&  18.86&  20.98\\
&85&   0.51& 100.75&   8.37&  28.97&  35.17&  19.03&  20.83\\
&86&   0.39& 107.39&   6.20&  28.75&  35.34&  19.25&  20.66\\
&87&   0.29& 112.67&   4.22&  28.57&  35.48&  19.43&  20.52\\
&88&   0.18& 116.02&   2.94&  28.44&  35.58&  19.56&  20.42\\
&89&   0& 118.13&   2.22&  28.34&  35.65&  19.66&  20.35\\
&90&   0& 119.51&   1.68&  28.27&  35.71&  19.73&  20.29\\
\colrule

150&87&   3.67&  81.96&  14.11&  28.28&  33.24&  19.72&  22.76\\
&88&   3.03&  82.38&  13.76&  28.16&  33.05&  19.84&  22.95\\
&89&   2.05&  83.21&  13.74&  28.34&  33.28&  19.66&  22.72\\
&90&   1.75&  86.33&  16.86&  28.45&  33.69&  19.55&  22.31\\
&91&   1.65&  86.04&  16.82&  28.79&  34.13&  19.21&  21.87\\
&92&   1.59&  84.08&  15.73&  28.92&  34.30&  19.08&  21.70\\
&93&   1.47&  80.67&  14.76&  28.82&  34.23&  19.18&  21.77\\
&94&   1.28&  78.84&  17.55&  28.92&  34.31&  19.08&  21.69\\
&95&   1.11&  85.49&  18.45&  29.33&  34.65&  18.67&  21.35\\
&96&   0.95&  89.87&  18.30&  29.45&  34.69&  18.55&  21.31\\
&97&   0.83&  94.63&  17.63&  29.42&  34.77&  18.58&  21.23\\
&98&   0.72& 100.59&  15.90&  29.33&  34.84&  18.67&  21.16\\
&99&   0.64& 106.35&  14.01&  29.23&  34.93&  18.77&  21.07\\
&100&   0.57& 111.73&  12.19&  29.14&  35.05&  18.86&  20.95\\
&101&   0.50& 117.18&  10.64&  29.03&  35.18&  18.97&  20.82\\
&102&   0.42& 123.24&   8.76&  28.88&  35.29&  19.12&  20.71\\
&103&   0.35& 129.13&   6.90&  28.73&  35.40&  19.27&  20.60\\
&104&   0.28& 133.97&   5.15&  28.59&  35.48&  19.41&  20.52\\
&105&   0.22& 137.47&   3.93&  28.47&  35.55&  19.53&  20.45\\
&106&   0.15& 140.00&   3.00&  28.38&  35.61&  19.62&  20.39\\
&107&   0.06& 141.92&   2.34&  28.31&  35.67&  19.69&  20.33\\
&108&   0& 143.25&   1.88&  28.25&  35.71&  19.75&  20.29\\
&109&   0& 144.25&   1.54&  28.21&  35.75&  19.79&  20.25\\
&110&   0& 145.01&   1.38&  28.18&  35.78&  19.82&  20.22
 \end{tabular}
 \end{ruledtabular}
 \end{table}

 \begin{table}[ht]
 %\caption{\label{tab:tip}}
 \begin{ruledtabular}
 \begin{tabular}{lcccc}
$E_{c.m.}$ & TKE$_\infty$ &  $L_{loss_\infty}$ & $Z_{H_\infty}$ & $N_{H_\infty}$\\
\colrule
150& 82.0 & 18.5 & 28.3& 33.2    \\
125 &  77.7 & 14.4  & &  \\
100 & 74.9 & 9.8 &   &  
  \end{tabular}
 \end{ruledtabular}
 \end{table}

\newpage
\subsection{$^{60}\mbox{Ni}+^{60}\mbox{Ni}$}

TDHF and TDRPA calculations for this system were published in \cite{williams2018}. 
They were performed by K. Sekizawa starting from his own TDHF code \cite{sekizawa2013} using the SLy4$d$ functional \cite{kim1997}.
For $E_{c.m.}=123$~MeV at $L_0=64$, TDRPA  give spurious results due to almost flat distributions near the TDHF average. 
The Coulomb capture barrier for this system is estimated to be $V_B=96.8$~MeV~\cite{swiatecki2005}.

  \begin{table}[ht]
% \caption{\label{tab:tip} }
 \begin{ruledtabular}
 \begin{tabular}{lccccc}
$E_{c.m.}$ & $L_0$ & $\tau$ & TKE & $\sigma_A$ & $\sigma_{NZ}$\\
\colrule
 110.4 &  47.3 &  2.68 &   92.5 &     3.028 &0.924  \\
     &  47.4 &  2.37 &   92.7 &     3.677 &1.483  \\
     &  48.1 &  1.83 &   91.0 &     2.034 &0.689  \\
     &  48.7 &  1.58 &   91.4 &     1.682 &0.678  \\
     &  49.3 &  1.46 &   91.5 &     1.548 &0.623  \\
     &  50.0 &  1.38 &   93.5 &     1.486 &0.582  \\
     &  50.6 &  1.28 &  106.2 &     1.675 &0.686  \\
     &  53.8 &  0.93 &  106.2 &     0.999 &0.225  \\
     &  53.3 &  0.87 &  107.4 &     0.854 &0.162  \\
     &  56.9 &  0.83 &  108.1 &     0.756 &0.126  \\
     &  60.1 &  0.77 &  108.8 &     0.618 &0.084  \\
     &  63.3 &  0.72 &  109.2 &     0.520 &0.058  \\
     &  75.9 &  0.51 &  109.8 &     0.281 &0.019  \\
     &  88.6 &  0.20 &  110.0 &     0.156 &0.009  \\
 \colrule
 123.0 &  64.0 &  3.52 &   94.6 &	      --  & --    \\
     &  64.1 &  2.97 &   96.4 &     6.632 &3.128  \\
     &  64.8 &  2.10 &   98.0 &     3.103 &1.289  \\
     &  65.4 &  1.80 &   94.5 &     2.737 &1.179  \\
     &  66.1 &  1.62 &   92.7 &     2.393 &1.019  \\
     &  66.8 &  1.47 &   94.7 &     1.986 &0.811  \\
     &  70.1 &  1.00 &  111.0 &     1.456 &0.496  \\
     &  73.4 &  0.81 &  119.1 &     0.908 &0.196  \\
     &  76.8 &  0.73 &  120.8 &     0.699 &0.113  \\
     &  80.1 &  0.68 &  121.5 &     0.568 &0.072  \\
     &  93.5 &  0.47 &  122.3 &     0.281 &0.018  \\
     & 106.8 &  0.19 &  122.5 &     0.147 &0.009  \\
     & 120.2 &  0 &  122.7 &     0.081 &0.007  \\
     & 133.5 &  0 &  122.7 &     0.051 &0.007  \\
\colrule
 135.6 &  71.9 &  4.10 &   98.2 &     7.533 &3.233  \\
     &  72.1 &  3.42 &   98.5 &     6.287 &2.040  \\
     &  72.2 &  3.16 &   97.7 &     7.428 &3.376  \\
     &  72.9 &  2.73 &   98.4 &     4.378 &1.996  \\
     &  73.6 &  2.51 &   99.4 &     4.089 &1.813  \\
     &  77.1 &  2.14 &  100.5 &     3.292 &1.373  \\
     &  80.6 &  1.37 &   99.7 &     2.301 &0.981  \\
     &  84.1 &  0.94 &  120.6 &     1.578 &0.555  \\
     &  87.6 &  0.76 &  130.7 &     0.974 &0.220  \\
     &  91.1 &  0.68 &  133.0 &     0.732 &0.121  \\
     &  98.1 &  0.57 &  134.3 &     0.473 &0.048  \\
     & 112.2 &  0.38 &  135.0 &     0.228 &0.014  \\
     & 126.2 &  0 &  135.2 &     0.117 &0.007  \\
     & 140.2 &  0 &  135.3 &     0.066 &0.007  
  \end{tabular}
 \end{ruledtabular}
 \end{table}

 \begin{table}[ht]
 %\caption{\label{tab:tip}}
 \begin{ruledtabular}
 \begin{tabular}{lccc}
$E_{c.m.}$ & TKE$_\infty$ &  $\sigma_{A_\infty}$ & $\sigma_{NZ_\infty}$\\
\colrule
135.6& 98.2 &   7.53 & 3.23    \\
123.0 & 94.6 &      &\\
110.4 & 92.5 &      &
  \end{tabular}
 \end{ruledtabular}
 \end{table}

\newpage
\subsection{$^{58}\mbox{Ni}+^{124}\mbox{Sn}$}

TDHF calculations for this system were published in \cite{wu2019}. 
They were performed by Wu and Guo with their modified version of the Sky3D code \cite{maruhn2014,schuetrumpf2018}
using the SLy5 functional \cite{chabanat1998a}.
The Coulomb capture barrier for this system is estimated to be $V_B=159.9$~MeV \cite{swiatecki2005}.

 \begin{table}[ht]
% \caption{\label{tab:tip} }
 \begin{ruledtabular}
 \begin{tabular}{lccccccc}
$E_{c.m.}$ & $L_0$ & $\tau$ & TKE & $Z_H$ & $N_H$ & $Z_L$ & $N_L$\\
\colrule
150&     0 &   0.53 &  139.8 &&&& \\
&     4.3 &   0.46 &  140.3 &&&&\\
&     8.5 &   0.47 &  141.6 &&&&\\
&    12.7 &   0.40 &  143.3&&&&\\
&    16.9 &   0.34 &  145.0&&&&\\
&    21.2 &   0.20 &  146.5 &&&&\\
&    25.4 &   0 &  147.5 &&&&\\
&    29.6 &   0 &  148.2 &&&&\\
&    33.8 &   0 &  148.8&&&&\\
\colrule
153&     0 &   1.66 &  137.1&&&&\\
&     4.3 &   1.66 &  136.8 &&&&\\
&     8.5 &   1.53 &  135.4 &&&&\\
&    12.8 &   1.33 &  133.2 &&&&\\
&    17.1 &   1.06 &  132.3&&&&\\
&    21.3 &   0.86 &  134.4 &&&&\\
&    25.7 &   0.67 &  138.9 &&&&\\
&    29.9 &   0.47 &  144.4 &&&&\\
&    34.1 &   0.26 &  148.3 &&&&\\
&    38.4 &   0.07 &  150.3 &&&&\\
&    42.7 &   0 &  151.3&&&&\\
&    46.9 &   0 &  151.9 &&&&\\
&    51.3 &   0 &  152.2 &&&&\\
\colrule
157&    29.6 &   3.60 &  130.3 &&&&\\
&    29.9 &   2.59 &  134.3&&&&\\
&    30.2 &   2.26 &  134.0 &&&&\\
&    31.1 &   1.87 &  137.8 &&&&\\
&    32.0 &   1.73 &  138.0 &&&&\\
&    34.6 &   1.33 &  135.1 &&&&\\
&    36.2 &   1.13 &  134.6 &&&&\\
&    38.9 &   0.86 &  137.2 &&&&\\
&    43.2 &   0.53 &  145.3 &&&&\\
&    47.5 &   0.33 &  151.7 &&&&\\
&    51.8 &   0.20 &  154.4 &&&&\\
&    60.4 &   0 &  156.0 &&&&\\
&    69.1 &   0 &  156.5 &&&&\\
\colrule
160.6&    42.5 &   3.66 &  132.5 &   51.3 &   70.1 &   26.7 &   33.9\\
&    42.9 &   2.40 &  134.7 &   50.9 &   69.6 &   27.1 &   34.4\\
&    43.7 &   1.80 &  137.3 &   50.8 &   70.3 &   27.2 &   33.7\\
&    45.4 &   1.40 &  136.7 &   51.3 &   71.4 &   26.7 &   32.6\\
&    47.2 &   1.13 &  136.1 &   51.7 &   72.2 &   26.3 &   31.8\\
&    48.0 &   0.99 &  136.6 &   51.9 &   72.5 &   26.1 &   31.5\\
&    49.9 &   0.86 &  139.5 &   52.1 &   72.8 &   25.9 &   31.2\\
&    52.5 &   0.66 &  145.6 &   51.8 &   73.1 &   26.2 &   30.9\\
&    56.8 &   0.40 &  154.5 &   50.8 &   73.4 &   27.2 &   30.6\\
&    61.1 &   0 &  157.8 &   50.4 &   73.6 &   27.6  &   30.4\\
&    65.5 &   0 &  159.1 &   50.2 &   73.7 &   27.8 &   30.3\\
&    70.0 &   0 &  159.6 &   50.1 &   73.8 &   27.9 &   30.2\\
  \end{tabular}
 \end{ruledtabular}
 \end{table}

Only the two highest energies lead to fusion and thus to fully damped collisions near the critical angular momentum for fusion. 
As the energies are relatively close, we do not expect a strong dependence of TKE$_\infty$ with the initial conditions. 
We thus fix TKE$_\infty\simeq133$~MeV. 
The neutron-proton equilibrium values are $Z_{H_\infty}\simeq51$ and $N_{H_\infty}\simeq70$.
\newpage

\newpage
\subsection{$^{54}\mbox{Cr}+^{186}\mbox{W}$} \label{sec:CrW}

The TDHF calculations for this system were performed by Umar and Oberacker.
They used their own code \cite{umar2006c} with the SLy4$d$ functional \cite{chabanat1998a}.
The $^{186}$W HF ground-state is found with a prolate deformation. 
The {\it tip} and {\it side} orientations, defined as deformation axis  respectively parallel and perpendicular to the collision axis have been studied.
The results for the side orientation have been published in \cite{hammerton2015}. 
The results for the tip orientation are unpublished. 
The Coulomb capture barrier for this system is estimated to be $V_B=192.6$~MeV \cite{swiatecki2005}.

 \begin{table}[ht]
% \caption{\label{tab:tip} }
 \begin{ruledtabular}
 \begin{tabular}{lccccccc}
$E_{c.m.}$ & Orientation & $L_0$ & $\tau$ & $Z_H$ & $A_H$ & $Z_L$ & $A_L$\\
\colrule
218.6& tip &    42.0 &   23.45 &   57.65 &  142.70 &   40.33 &   97.28\\
& &    50.4 &   20.36 &   56.53 &  140.25 &   41.47 &   99.74\\
& &    56.7 &   14.20 &   55.83 &  137.54 &   42.14 &  102.40\\
& &    59.9 &   13.09 &   58.68 &  145.70 &   39.31 &   94.30\\
& &    63.0 &   12.00 &   60.69 &  150.31 &   37.30 &   89.65\\
& &    84.0 &    6.24 &   66.98 &  165.78 &   31.02 &   74.21\\
& &   105.0 &    1.73 &   73.80 &  183.02 &   24.15 &   56.95\\
& side  &    52.5 &   18.30 &   53.90 &  132.74 &   44.10 &  107.26\\
& &    54.6 &   13.60 &   55.26 &  136.17 &   42.73 &  103.81\\
& &    56.7 &   12.30 &   55.78 &  137.55 &   42.21 &  102.42\\
& &    59.9 &   11.90 &   56.42 &  139.79 &   41.56 &  100.18\\
& &    63.0 &   11.80 &   56.28 &  139.74 &   41.71 &  100.24\\
& &    73.5 &    9.74 &   63.38 &  156.20 &   34.60 &   83.77\\
& &    77.7 &    9.35 &   62.63 &  154.96 &   35.36 &   85.04\\
& &    84.0 &    6.71 &   66.83 &  165.29 &   31.16 &   74.67\\
& &   105.0 &    2.45 &   73.09 &  181.04 &   24.89 &   58.93
  \end{tabular}
 \end{ruledtabular}
 \end{table}

\newpage
\subsection{$^{40}\mbox{Ca}+^{208}\mbox{Pb}$}

TDHF calculations for this system have been made by Simenel with the TDHF3D code and the SLy4$d$ functional \cite{kim1997}.
The results were published in Ref.~\cite{simenel2012b}.
The Coulomb capture  barrier for this system is estimated to be $V_B=179.5$~MeV \cite{swiatecki2005}.

 \begin{table}[ht]
% \caption{\label{tab:tip} }
 \begin{ruledtabular}
 \begin{tabular}{lccccccc}
$E_{c.m.}$ & $L_0$ & $\tau$ & $L_{loss}$ & $Z_H$ & $N_H$ & $Z_L$ & $N_L$\\
\colrule
179.1&20&   1.95&   2.52&  83.89& 122.67&  18.11&  23.33\\
&22&   1.43&   3.49&  83.74& 122.89&  18.26&  23.11\\
&24&   1.20&   4.59&  83.65& 123.07&  18.35&  22.93\\
&26&   1.06&   4.61&  83.50& 123.35&  18.50&  22.65\\
&28&   0.95&   4.05&  83.36& 123.57&  18.64&  22.43\\
&30&   0.85&   5.06&  83.40& 123.81&  18.60&  22.19\\
&32&   0.76&   3.85&  83.62& 124.05&  18.38&  21.95\\
&34&   0.67&   5.66&  83.71& 124.17&  18.29&  21.83\\
&36&   0.60&   4.22&  83.63& 124.32&  18.37&  21.68\\
&38&   0.53&   4.61&  83.47& 124.51&  18.53&  21.49\\
&40&   0.47&   2.68&  83.21& 124.73&  18.79&  21.27\\
&42&   0.41&   2.04&  82.93& 124.92&  19.07&  21.08\\
&44&   0.36&   0.80&  82.70& 125.12&  19.30&  20.88\\
&46&   0.33&   1.30&  82.55& 125.27&  19.45&  20.73\\
&48&   0.30&   1.14&  82.44& 125.39&  19.56&  20.61\\
&50&   0.28&   0.46&  82.36& 125.47&  19.64&  20.53\\
&52&   0.27&   0.53&  82.30& 125.54&  19.70&  20.46\\
&54&   0.25&   0.11&  82.25& 125.60&  19.75&  20.40\\
%&56&   0.24&  -1.09&  82.21& 125.65&  19.79&  20.35\\
%&58&   0.22&   0.97&  82.18& 125.69&  19.82&  20.31\\
%&60&   0.21&   0.75&  82.15& 125.72&  19.85&  20.28\\
%&70&   0.15&  -0.33&  82.07& 125.84&  19.93&  20.16
  \end{tabular}
 \end{ruledtabular}
 \end{table}

 The values for neutron-proton equilibrium are $Z_{H_\infty}\simeq83.9$ and $N_{H_\infty}\simeq122.7$.
 For angular momentum dissipation, we take $L_{loss_\infty}\simeq5.7$. 
 
 \newpage
\subsection{$^{78}\mbox{Kr}+^{208}\mbox{Pb}$}

TDHF results for this system have been published in \cite{umar2017}.
The calculations were made by Umar using his  code \cite{umar2006c}  with the SLy4$d$ functional \cite{kim1997}.
%The $^{78}$Kr HF ground-state is found with a prolate deformation ($\beta_2\simeq0.088$). 
The orientation of $^{78}$Kr is defined as in Sec.~\ref{sec:CrW}. The Coulomb capture  barrier for this system is estimated to be $V_B=306.7$~MeV \cite{swiatecki2005}.

  \begin{table}[ht]
% \caption{\label{tab:tip} }
 \begin{ruledtabular}
 \begin{tabular}{lccccccccc}
$E_{c.m.}$ & Orientation & $L_0$ & $\tau$ & TKE & $L_{loss}$ & $Z_H$ & $N_H$ & $Z_L$ & $N_L$\\
\colrule
482 &tip&     0 &    3.04 &   233.4 &     0.0 &    84.2 &   122.2 &    33.8 &    45.8\\
&&    18.1 &    2.85 &   237.5 &     3.0 &    84.0 &   122.0 &    33.9 &    46.0\\
&&    36.3 &    2.99 &   236.6 &     8.0 &    83.5 &   121.0 &    34.5 &    47.0\\
&&    54.5 &    3.23 &   234.9 &    16.0 &    83.5 &   120.4 &    34.5 &    47.6\\
&&    77.6 &    3.25 &   236.7 &    30.0 &    82.9 &   119.6 &    35.1 &    48.4\\
&&    98.8 &    3.55 &   228.2 &    31.0 &    81.5 &   117.9 &    36.5 &    50.1\\
&&   108.9 &    3.71 &   234.1 &    23.0 &    79.4 &   114.5 &    38.6 &    53.4\\
&&   119.8 &    3.49 &   227.5 &    44.0 &    80.5 &   116.3 &    37.4 &    51.7\\
&&   127.1 &    3.49 &   233.3 &    24.0 &    80.9 &   117.3 &    37.1 &    50.7\\
&&   145.2 &    3.15 &   239.7 &    22.0 &    80.8 &   116.8 &    37.2 &    51.2\\
&&   163.4 &    2.99 &   246.9 &    33.0 &    80.5 &   116.2 &    37.5 &    51.8\\
&&   181.6 &    2.99 &   248.4 &    30.0 &    79.5 &   114.7 &    38.5 &    53.3\\
&&   199.7 &    2.80 &   243.2 &    32.0 &    79.9 &   115.5 &    38.1 &    52.5\\
&&   217.9 &    2.91 &   241.1 &    50.0 &    79.4 &   114.7 &    38.6 &    53.3\\
&&   236.0 &    1.68 &   264.3 &    60.0 &    82.6 &   119.6 &    35.4 &    48.4\\
&&   254.2 &    1.33 &   277.9 &    71.0 &    83.9 &   121.4 &    34.1 &    46.6\\
&&   272.3 &    0.93 &   328.0 &    56.0 &    84.9 &   123.1 &    33.1 &    44.9\\
&&   290.5 &    0.59 &   391.9 &    33.0 &    83.8 &   124.5 &    34.2 &    43.5\\
&&   308.6 &    0.45 &   438.0 &    29.0 &    82.8 &   124.6 &    35.2 &    43.4\\
&&   326.8 &    0.29 &   444.2 &    24.0 &    82.7 &   125.4 &    35.2 &    42.6\\
&&   354.0 &    0.19 &   472.8 &    14.0 &    82.1 &   125.8 &    35.9 &    42.2\\
&side&     0 &    4.53 &   236.2 &     0.0 &    78.6 &   113.4 &    39.4 &    54.6\\
&&    36.3 &    4.67 &   260.3 &     6.2 &    77.1 &   111.5 &    40.8 &    56.5\\
&&    77.6 &    5.07 &   248.5 &    18.2 &    78.7 &   113.5 &    39.3 &    54.5\\
&&   108.9 &    4.27 &   261.5 &    21.1 &    78.5 &   113.6 &    39.5 &    54.4\\
&&   145.2 &    4.00 &   238.6 &    17.5 &    78.3 &   113.2 &    39.7 &    54.8\\
&&   181.6 &    3.23 &   251.7 &    30.0 &    78.4 &   113.3 &    39.6 &    54.7\\
&&   217.9 &    2.67 &   235.0 &    57.7 &    81.4 &   117.7 &    36.6 &    50.3\\
&&   254.2 &    1.49 &   283.4 &    68.0 &    84.4 &   122.7 &    33.6 &    45.3\\
&&   290.5 &    0.85 &   349.0 &    50.8 &    84.8 &   124.0 &    33.2 &    44.0\\
&&   326.8 &    0.43 &   451.3 &    12.6 &    82.8 &   124.9 &    35.2 &    43.0
  \end{tabular}
 \end{ruledtabular}
 \end{table}

The neutron-to-proton equilibrium values are chosen before a significant mass drift occurs for the most central collisions. 

 \begin{table}[ht]
 %\caption{\label{tab:tip}}
 \begin{ruledtabular}
 \begin{tabular}{lcccc}
$E_{c.m.}$  & TKE$_\infty$ &  $L_{loss_\infty}$ & $Z_{H_\infty}$ & $N_{H_\infty}$\\
\colrule
482 &228 & 71 & 79.4 & 114.7    \\
  \end{tabular}
 \end{ruledtabular}
 \end{table}

 \newpage
\subsection{$^{92}\mbox{Kr}+^{208}\mbox{Pb}$}

TDHF results for this system have been published in \cite{umar2017}.
The calculations were made by Umar using his  code \cite{umar2006c}  with the SLy4$d$ functional \cite{kim1997}.
The orientation of $^{92}$Kr is defined as in Sec.~\ref{sec:CrW}. 
%The $^{92}$Kr HF ground-state is found with a prolate deformation ($\beta_2\simeq0.178$). 
%The {\it tip} and {\it side} orientations, defined as deformation axis  respectively parallel and perpendicular to the collision axis have been studied.
The Coulomb capture  barrier for this system is estimated to be $V_B=299.5$~MeV \cite{swiatecki2005}.

  \begin{table}[ht]
% \caption{\label{tab:tip} }
 \begin{ruledtabular}
 \begin{tabular}{lccccccccc}
$E_{c.m.}$ & Orientation & $L_0$ & $\tau$ & TKE & $L_{loss}$ & $Z_H$ & $N_H$ & $Z_L$ & $N_L$\\
\colrule
542&tip&     0.0 &    4.05 &   225.2 &     0.0 &    77.5 &   120.9 &    40.5 &    61.1\\
&&    20.4 &    3.87 &   230.4 &     1.9 &    78.3 &   121.9 &    39.7 &    60.1\\
&&    40.8 &    3.92 &   232.6 &     8.7 &    77.6 &   121.0 &    40.4 &    61.0\\
&&    61.2 &    3.92 &   240.0 &    13.0 &    76.2 &   118.9 &    41.8 &    63.1\\
&&    81.6 &    3.81 &   253.7 &    27.3 &    74.9 &   117.1 &    43.1 &    64.9\\
&&   102.0 &    3.89 &   256.2 &    28.2 &    74.7 &   116.3 &    43.3 &    65.6\\
&&   122.5 &    3.95 &   255.6 &    29.3 &    74.8 &   116.6 &    43.1 &    65.3\\
&&   142.9 &    4.00 &   246.5 &    23.9 &    76.3 &   119.1 &    41.6 &    62.9\\
&&   163.3 &    3.76 &   236.5 &    34.6 &    79.0 &   123.4 &    38.9 &    58.6\\
&&   183.7 &    3.65 &   231.3 &    46.1 &    81.3 &   127.2 &    36.7 &    54.8\\
&&   204.1 &    3.62 &   226.3 &    45.1 &    80.9 &   126.7 &    37.1 &    55.3\\
&&   224.5 &    3.47 &   227.4 &    46.9 &    80.0 &   125.5 &    38.0 &    56.5\\
&&   245.0 &    3.41 &   243.8 &    56.0 &    78.7 &   122.9 &    39.3 &    59.1\\
&&   265.4 &    2.77 &   270.5 &    57.0 &    75.4 &   117.5 &    42.5 &    64.5\\
&&   285.8 &    2.19 &   281.7 &    60.0 &    77.1 &   120.4 &    40.9 &    61.6\\
&&   306.2 &    1.68 &   296.5 &    90.1 &    78.4 &   123.1 &    39.6 &    58.9\\
&&   326.7 &    1.25 &   323.6 &    99.3 &    79.9 &   125.1 &    38.1 &    56.9\\
&&   347.0 &    1.04 &   356.0 &    96.0 &    80.8 &   125.8 &    37.2 &    56.2\\
&&   367.5 &    0.69 &   451.9 &    47.8 &    81.5 &   126.0 &    36.5 &    56.0\\
&&   388.0 &    0.37 &   514.0 &    17.8 &    81.8 &   126.1 &    36.2 &    55.9\\
&side&     0.0 &    6.56 &   258.4 &     0.0 &    75.5 &   117.7 &    42.5 &    64.3\\
&&    40.8 &    6.72 &   248.0 &     7.9 &    77.1 &   120.4 &    40.9 &    61.6\\
&&    81.6 &    6.40 &   244.4 &    17.7 &    75.9 &   119.3 &    42.1 &    62.7\\
&&   122.5 &    6.15 &   245.5 &     5.9 &    75.0 &   118.4 &    43.0 &    63.6\\
&&   163.3 &    4.99 &   258.8 &    37.0 &    75.0 &   117.2 &    43.0 &    64.8\\
&&   204.1 &    3.84 &   259.8 &    34.1 &    74.4 &   116.7 &    43.5 &    65.3\\
&&   245.0 &    3.20 &   246.4 &    42.3 &    77.0 &   120.1 &    41.0 &    61.9\\
&&   285.8 &    2.35 &   253.5 &    79.1 &    79.7 &   125.0 &    38.3 &    57.0\\
&&   326.7 &    1.65 &   286.9 &    93.7 &    80.5 &   126.3 &    37.5 &    55.7\\
&&   367.5 &    1.00 &   370.5 &    81.2 &    81.4 &   127.5 &    36.6 &    54.5\\
&&   408.3 &    0.63 &   462.4 &    42.8 &    81.9 &   127.4 &    36.1 &    54.6
  \end{tabular}
 \end{ruledtabular}
 \end{table}

 The estimated values for energy and angular momentum dissipation are TKE$_\infty\simeq240$ MeV and $L_{loss_\infty}\simeq99$. 
 
 \newpage
\subsection{$^{176}\mbox{Yb}+^{176}\mbox{Yb}$}

TDHF and TDRPA 
calculations for this system have been reported in \cite{godbey2020b} with the code of \cite{umar2006c} and the SLy4$d$ functional \cite{kim1997}. The angular momentum loss has been computed for the present work.  
The orientation is defined as in Sec.~\ref{sec:CrW}. Both collision partners have the same initial orientation. 
The Coulomb capture barrier for this system is estimated to be $V_B=446.5$~MeV \cite{swiatecki2005}.
For tip orientation at $E_{c.m.}=660$~MeV and $L_0=0$,  TDRPA  gives spurious results due to almost flat distributions near the TDHF average \cite{godbey2020b}. Negative values of $L_{loss}$ indicate angular momentum transfer from the fragments  to the orbital motion. 
Equilibrium values are TKE$_\infty\simeq310$~MeV, $L_{loss_\infty}\simeq94$ (at 660 MeV), $L_{loss_\infty}\simeq149$ (at 880 MeV), $\sigma_{A_\infty}\simeq25$, and $\sigma_{NZ_\infty}\simeq13$.
  \begin{table}[ht]
% \caption{\label{tab:tip} }
 \begin{ruledtabular}
 \begin{tabular}{lccccccc}
$E_{c.m.}$ & Orientation & $L_0$ & $\tau$ & TKE & $L_{loss}$ & $\sigma_A$ & $\sigma_{NZ}$\\
\colrule
660&tip&     0 &    1.96 &   340.7 &   0 &  -- &   --\\
&&    52.9 &    3.69 &   269.2 &    -3.4 &   30.83 &   14.71\\
&&   105.8 &    3.39 &   308.5 &     7.6 &   23.27 &   10.38\\
&&   158.7 &    4.17 &   279.2 &    38.4 &   26.59 &   13.01\\
&&   211.7 &    4.16 &   291.8 &    31.3 &   23.01 &   11.11\\
&&   264.6 &    3.23 &   302.1 &    48.6 &   46.03 &   22.76\\
&&   317.5 &    1.73 &   405.5 &    93.2 &   13.60 &    6.54\\
&&   370.4 &    1.17 &   480.5 &    94.3 &    6.77 &    3.17\\
&&   423.3 &    0.45 &   608.0 &    24.0 &    2.46 &    0.94\\
&&   476.2 &    0 &   653.8 &   -13.8 &    0.64 &    0.07\\
&side&     0 &    3.06 &   399.6 &    -0.6 &    8.35 &    3.79\\
&&    52.9 &    3.55 &   389.0 &    14.6 &   20.41 &    9.35\\
&&   105.8 &    3.87 &   355.6 &    21.0 &   49.42 &   23.38\\
&&   158.7 &    3.29 &   358.4 &    18.5 &   23.23 &   10.93\\
&&   211.7 &    2.76 &   368.6 &    31.6 &   25.02 &   12.05\\
&&   264.6 &    2.17 &   376.4 &    33.9 &   18.10 &    8.67\\
&&   317.5 &    1.73 &   405.3 &    47.7 &   14.11 &    6.74\\
&&   370.4 &    1.78 &   394.5 &    49.7 &   40.90 &   20.08\\
&&   423.3 &    1.31 &   449.4 &    74.7 &    9.85 &    4.72\\
&&   476.2 &    0.91 &   516.0 &    69.9 &    5.48 &    2.54\\
&&   529.1 &    0.39 &   607.5 &    45.2 &    2.24 &    0.88\\
&&   582.1 &    0 &   653.5 &    22.3 &    0.58 &    0.07\\
%&&   635.0 &    0.00 &   656.5 &    18.5 &    0.28 &    0.01\\
%&&   687.9 &    0.00 &   657.6 &    14.6 &    0.14 &    0.00\\
%&&   740.8 &    0.00 &   658.3 &    11.9 &    0.08 &    0.00\\
\colrule
880&tip&     0 &    1.18 &   394.1 &    0 &   25.64 &   12.52\\
&&    61.1 &    1.25 &   377.3 &    50.2 &   20.69 &    9.52\\
&&   122.2 &    2.87 &   309.1 &    15.7 &   79.80 &   39.68\\
&&   183.3 &    3.61 &   298.1 &    48.2 &   30.65 &   13.20\\
&&   244.4 &    4.14 &   312.4 &    46.0 &   32.60 &   15.11\\
&&   305.5 &    3.54 &   316.4 &    74.3 &   37.14 &   17.67\\
&&   366.6 &    2.65 &   359.9 &    97.5 &   23.08 &   11.17\\
&&   427.7 &    1.83 &   430.5 &   119.7 &   21.40 &   10.26\\
&&   488.8 &    1.15 &   526.1 &   148.6 &   10.43 &    4.97\\
&&   549.9 &    0.63 &   696.8 &    82.6 &    5.01 &    2.20\\
&&   611.0 &    0 &   860.4 &     3.0 &    1.18 &    0.28\\
%&&   672.1 &    0.00 &   875.3 &    -6.0 &    0.47 &    0.03\\
&side&     0 &    2.57 &   401.0 &    -0.2 &   10.55 &    4.78\\
&&    61.1 &    2.30 &   409.8 &    14.0 &   11.43 &    4.71\\
&&   122.2 &    2.53 &   396.2 &    35.3 &   14.64 &    6.98\\
&&   183.3 &    2.82 &   377.9 &    48.9 &   17.93 &    7.93\\
&&   244.4 &    2.95 &   388.9 &    40.0 &   46.75 &   22.73\\
&&   305.5 &    2.84 &   361.1 &    60.2 &   51.42 &   25.12\\
&&   366.6 &    2.46 &   381.9 &    62.1 &   44.32 &   21.44\\
&&   427.7 &    2.19 &   389.4 &    69.5 &   64.43 &   31.46\\
&&   488.8 &    1.84 &   425.0 &    63.2 &   36.69 &   17.85\\
&&   549.9 &    1.65 &   466.0 &    72.9 &   38.86 &   19.09\\
&&   611.0 &    1.20 &   520.5 &   118.2 &   15.62 &    7.53\\
&&   672.1 &    0.78 &   644.1 &   117.1 &    6.62 &    3.11\\
&&   733.2 &    0.28 &   820.1 &    38.2 &    2.17 &    0.78\\
&&   794.3 &    0 &   873.1 &    14.5 &    0.61 &    0.06\\
%&&   855.4 &    0.00 &   876.7 &    11.6 &    0.28 &    0.01
  \end{tabular}
 \end{ruledtabular}
 \end{table}
 
  \newpage
\subsection{$^{40}\mbox{Ca}+^{238}\mbox{U}$}

TDHF calculations for this system have been reported by Wakhle {\it et al.} \cite{wakhle2014} with the TDHF3D code \cite{kim1997} and by Oberacker {\it et al.}  \cite{oberacker2014} with the code of \cite{umar2006c}.
Both studies used the SLy4$d$ functional \cite{kim1997}.
The orientation is defined as in Sec.~\ref{sec:CrW}. 
The Coulomb capture barrier for this system is estimated to be $V_B=197.3$~MeV \cite{swiatecki2005}.
 \begin{table}[ht]
% \caption{\label{tab:tip} }
 \begin{ruledtabular}
 \begin{tabular}{lcccccccc}
Author&$E_{c.m.}$ & Orientation & $L_0$ & $\tau$ & $Z_H$ & $A_H$ & $Z_L$ & $A_L$\\
\colrule
Oberacker et al	&	200.0	&	side	&	0	&	1.29	&	94.7	&	236.8	&	17.3	&	41.2	\\
	&	203.0	&		&		&	3.66	&	92.4	&	231.3	&	19.6	&	46.7	\\
	&	204.0	&		&		&	4.75	&	90.7	&	226.6	&	21.3	&	51.4	\\
	&	205.0	&		&		&	5.02	&	90.8	&	227.0	&	21.2	&	51.0	\\
	&	206.0	&		&		&	7.35	&	88.8	&	222.2	&	23.2	&	55.8	\\
	&	207.0	&		&		&	12.80	&	82.8	&	206.4	&	29.2	&	71.6	\\
	&	208.0	&		&		&	21.71	&	71.8	&	179.5	&	40.2	&	98.5	\\
	&	209.0	&		&		&	20.67	&	70.8	&	177.0	&	41.2	&	101.0	\\
	&	210.0	&		&		&	21.81	&	69.8	&	174.8	&	42.2	&	103.2	\\
	&	211.0	&		&		&	18.06	&	72.5	&	181.2	&	39.5	&	96.8	\\
	&	215.0	&		&		&	19.39	&	69.2	&	172.7	&	42.8	&	105.3	\\
	&	220.0	&		&		&	32.24	&	68.7	&	170.9	&	43.3	&	107.1	\\
\colrule
Wakhle et al	&	186.6	&	tip	&	0	&	6.30	&	81.0	&	203.0	&	31.0	&	75.0	\\
	&		&		&	10	&	5.91	&	81.6	&	204.8	&	30.4	&	73.2	\\
	&		&		&	20	&	6.58	&	80.7	&	203.0	&	31.3	&	75.0	\\
	&		&		&	30	&	6.06	&	86.2	&	216.4	&	25.8	&	61.6	\\
	&		&		&	35	&	8.50	&	85.9	&	216.3	&	26.1	&	61.7	\\
	&		&		&	40	&	9.36	&	84.8	&	213.0	&	27.2	&	65.0	\\
	&	&		&	45	&	7.30	&	87.6	&	220.1	&	24.4	&	57.9	\\
	&		&		&	50	&	3.31	&	88.4	&	222.4	&	23.6	&	55.6	\\
	&		&		&	60	&	1.93	&	93.3	&	237.9	&	18.7	&	40.1	\\
\colrule
Wakhle et al	&	205.9	&	tip	&	0	&	8.40	&	81.6	&	204.1	&	30.4	&	73.9	\\
	&		&		&	10	&	7.55	&	81.0	&	204.0	&	31.0	&	74.0	\\
	&		&		&	20	&	7.37	&	81.9	&	204.9	&	30.1	&	73.1	\\
	&		&		&	30	&	7.19	&	81.0	&	202.8	&	31.0	&	75.2	\\
	&		&		&	40	&	6.29	&	81.0	&	204.0	&	31.0	&	74.0	\\
	&		&		&	50	&	6.53	&	80.0	&	201.9	&	32.0	&	76.1	\\
	&		&		&	60	&	6.25	&	80.0	&	201.0	&	32.0	&	77.0	\\
	&		&		&	70	&	6.10	&	80.0	&	201.9	&	32.0	&	76.1	\\
	&		&		&	80	&	6.24	&	86.9	&	217.9	&	25.1	&	60.1	\\
	&		&		&	90	&	3.53	&	91.9	&	231.0	&	20.1	&	47.0	\\
	&		&		&	100	&	1.86	&	91.9	&	237.9	&	20.1	&	40.1	\\
	&		&	side	&	20	&	11.89	&	67.0	&	167.0	&	45.0	&	111.0	\\
	&		&		&	30	&	9.51	&	80.0	&	202.1	&	32.0	&	75.9	\\
	&		&		&	40	&	5.87	&	94.0	&	235.9	&	18.0	&	42.1	\\
	&		&		&	50	&	1.45	&	93.0	&	237.0	&	19.0	&	41.0	\\
	&		&		&	100	&	0.50	&	92.1	&	238.2	&	19.9	&	39.8	\\
\colrule
Wakhle et al	&	225.4	&	tip	&	0	&	6.46	&	81.6	&	204.1	&	30.4	&	73.9	\\
	&		&		&	10	&	6.43	&	81.6	&	205.0	&	30.4	&	73.1	\\
	&		&		&	20	&	6.21	&	82.1	&	206.1	&	29.9	&	71.9	\\
	&		&		&	30	&	5.86	&	83.1	&	208.1	&	28.9	&	69.9	\\
	&		&		&	40	&	5.85	&	83.0	&	208.0	&	29.0	&	70.0	\\
	&		&		&	50	&	5.75	&	81.1	&	202.9	&	30.9	&	75.1	\\
	&		&		&	60	&	5.80	&	81.0	&	202.1	&	31.0	&	75.9	\\
	&		&		&	70	&	5.91	&	80.0	&	201.0	&	32.0	&	77.0	\\
	&		&		&	80	&	5.91	&	81.0	&	201.9	&	31.0	&	76.1	\\
	&		&		&	90	&	5.88	&	80.1	&	201.1	&	31.9	&	76.9	\\
	&		&		&	100	&	5.73	&	80.1	&	201.1	&	31.9	&	76.9	\\
	&		&		&	110	&	4.83	&	90.0	&	225.0	&	22.0	&	53.0	\\
	&		&		&	120	&	2.79	&	92.0	&	234.0	&	20.0	&	44.0	\\
	&		&	side	&	40	&	29.08	&	56.0	&	140.1	&	56.0	&	137.9	\\
	&		&		&	50	&	13.07	&	66.0	&	166.1	&	46.0	&	111.9	\\
	&		&		&	55	&	11.43	&	57.9	&	143.8	&	54.1	&	134.3	\\
	&		&		&	60	&	9.92	&	60.9	&	142.9	&	51.1	&	135.1	\\
	&		&		&	65	&	11.49	&	66.0	&	166.1	&	46.0	&	111.9	\\
	&		&		&	70	&	8.24	&	82.0	&	205.0	&	30.0	&	73.0	\\
	&		&		&	75	&	4.78	&	90.0	&	224.9	&	22.0	&	53.1	\\
	&		&		&	80	&	1.82	&	94.0	&	238.0	&	18.0	&	40.0	\\
	&		&		&	100	&	0.74	&	92.0	&	238.1	&	20.0	&	39.9	
  \end{tabular}
 \end{ruledtabular}
 \end{table}

\newpage
\subsection{$^{50}\mbox{Ti}+^{249}\mbox{Bk}$}
TDHF calculations for this system have been reported  in \cite{umar2016} using the code of \cite{umar2006c} with the SLy4$d$ functional \cite{kim1997}.
The orientation is defined as in Sec.~\ref{sec:CrW}. 
The Coulomb capture barrier for this system is estimated to be $V_B=222.8$~MeV \cite{swiatecki2005}.
 \begin{table}[ht]
\begin{ruledtabular}
 \begin{tabular}{lccccccc}
$E_{c.m.}$ & Orientation & $L_0$ & $\tau$   & $Z_H$ & $A_H$ & $Z_L$ & $A_L$\\
\colrule
207.5	&	tip	&	0	&	1.37	&	97.5	&	248.6	&	21.5	&	50.4	\\
214.8	&		&		&	15.52	&	81.4	&	207.0	&	37.6	&	92.0	\\
220.6	&		&		&	12.51	&	79.9	&	203.1	&	39.1	&	95.9	\\
233.0	&		&		&	12.34	&	81.6	&	208.8	&	37.4	&	90.2	\\
233.2	&		&		&	12.55	&	81.6	&	208.8	&	37.4	&	90.2	\\
	&		&	21.6	&	8.51	&	81.9	&	208.0	&	37.1	&	90.9	\\
	&		&	43.3	&	7.45	&	81.5	&	206.8	&	37.5	&	92.2	\\
	&		&	64.9	&	8.09	&	82.2	&	208.1	&	36.8	&	90.9	\\
245.8	&		&	0	&	21.77	&	82.9	&	210.6	&	36.1	&	88.4	\\
\colrule
223.9	&	side	&	0	&	0.92	&	98.1	&	249.0	&	20.9	&	50.0	\\
225.9	&		&		&	1.55	&	97.6	&	247.7	&	21.4	&	51.3	\\
226.9	&		&		&	2.61	&	97.0	&	246.6	&	22.0	&	52.4	\\
227.6	&		&		&	7.25	&	94.5	&	240.2	&	24.5	&	58.8	\\
233.2	&		&	21.6	&	15.96	&	75.7	&	192.2	&	43.2	&	106.8	\\
	&		&	43.3	&	8.40	&	77.9	&	197.6	&	41.1	&	101.3	\\
	&		&	64.9	&	1.17	&	97.9	&	249.0	&	21.1	&	50.0	   
\end{tabular}
 \end{ruledtabular}
 \end{table}

\newpage
\subsection{$^{48}\mbox{Ca}+^{249}\mbox{Bk}$}\label{sec:CaBk}

TDHF calculations for this system have been reported  in \cite{umar2016,godbey2019} using the code of \cite{umar2006c} with the SLy4$d$ functional \cite{kim1997}.
The orientation is defined by the angle (in degrees) between the deformation axis and the collision axis. 
This is the system with the largest number of data so the results are presented in 4 tables. 
TKE was not computed at $E_{c.m.}=218$~MeV. 
The Coulomb capture barrier for this system is estimated to be $V_B=202.0$~MeV \cite{swiatecki2005}. The TKE dissipation is obtained before significant mass transfer occurs, leading to TKE$_\infty\simeq170$~MeV. 
  \begin{table}[ht]
% \caption{\label{tab:tip} }
 \begin{ruledtabular}
 \begin{tabular}{lcccccccc}
$E_{c.m.}$ & Orientation & $L_0$ & $\tau$ & TKE  & $Z_H$ & $N_H$ & $Z_L$ & $N_L$\\
\colrule
209& 0 & 0 &    9.74 &   223.1 &    79.6 &   125.1 &    37.4 &    54.9\\
&  & 10 &    8.14 &   237.2 &    78.7 &   123.0 &    38.3 &    57.0\\
&  & 20 &    7.27 &   243.7 &    78.0 &   121.0 &    39.0 &    59.0\\
&  & 30 &    6.34 &   227.7 &    80.2 &   125.1 &    36.8 &    54.9\\
&  & 40 &    5.70 &   234.7 &    80.1 &   125.1 &    36.9 &    54.9\\
&  & 50 &    5.70 &   234.8 &    80.2 &   125.4 &    36.8 &    54.6\\
&  & 60 &    7.00 &   232.8 &    77.0 &   119.7 &    40.0 &    60.3\\
&  & 70 &    6.64 &   234.3 &    78.7 &   122.3 &    38.3 &    57.7\\
&  & 80 &    6.20 &   196.9 &    83.0 &   130.3 &    34.0 &    49.7\\
&  & 90 &    1.23 &   171.9 &    96.4 &   150.8 &    20.6 &    29.2\\
& 30 & 0 &    5.47 &   234.7 &    80.6 &   126.2 &    36.4 &    53.8\\
& & 10 &    6.04 &   238.0 &    79.5 &   124.5 &    37.4 &    55.5\\
&  & 20 &    7.00 &   230.4 &    78.4 &   122.6 &    38.6 &    57.4\\
&  & 30 &    6.90 &   233.8 &    77.8 &   121.6 &    39.2 &    58.4\\
&  & 40 &    7.37 &   210.1 &    81.2 &   127.0 &    35.8 &    53.0\\
&  & 50 &    2.70 &   168.9 &    94.9 &   149.1 &    22.1 &    30.9\\
& 45 & 0 &    6.67 &   262.1 &    79.5 &   137.2 &    37.5 &    62.8\\
&  & 10 &    7.54 &   225.9 &    78.9 &   122.9 &    38.1 &    57.0\\
&  & 20 &    8.57 &   195.5 &    80.5 &   126.2 &    41.6 &    61.3\\
&  & 30 &    2.87 &   169.8 &    94.5 &   148.3 &    22.5 &    31.7\\
&  & 40 &    0.50 &   188.2 &    97.0 &   151.8 &    20.0 &    28.2\\
& 60 & 0 &    8.67 &   206.4 &    76.5 &   119.8 &    40.5 &    60.2\\
&  & 10 &    8.54 &   210.3 &    80.4 &   126.1 &    36.5 &    53.9\\
&  & 20 &    6.80 &   178.7 &    84.5 &   132.8 &    32.5 &    47.2\\
&  & 30 &    2.50 &   174.9 &    94.9 &   149.0 &    22.1 &    31.0\\
&  & 40 &    0.77 &   186.2 &    96.8 &   151.2 &    20.2 &    28.8\\
& 90 & 0 &   14.54 &   201.5 &    85.7 &   133.7 &    31.3 &    46.3\\
&  & 10 &    5.44 &   180.5 &    88.8 &   139.4 &    28.2 &    40.6\\
&  & 20 &    7.30 &   176.5 &    88.3 &   139.0 &    28.7 &    41.0\\
&  & 30 &    4.27 &   162.2 &    92.4 &   145.0 &    24.6 &    35.0\\
&  & 40 &    1.93 &   169.3 &    95.7 &   149.6 &    21.3 &    30.4\\
\colrule
218& 0 &     0&   12.00 &  -- &    77.5 &   120.5 &    39.5 &    59.4\\
&  &    10.3 &    9.27 &  -- &    76.8 &   119.1 &    40.1 &    60.9\\
&  &    20.6 &    7.92 &  -- &    78.1 &   121.3 &    38.9 &    58.7\\
&  &    41.1 &    6.12 &  -- &    79.5 &   124.3 &    37.5 &    55.7\\
&  &    61.7 &    6.57 &  -- &    79.9 &   124.6 &    37.1 &    55.3\\
&  &    82.3 &    6.75 &  -- &    83.5 &   131.1 &    33.5 &    49.0\\
&  &   102.8 &    0.63 &  -- &    97.1 &   151.8 &    19.9 &    28.2\\
& 90 &     6.2 &   12.70 &  -- &    74.1 &   115.7 &    42.9 &    64.1\\
&  &    10.3 &   13.60 & -- &    73.4 &   114.2 &    43.6 &    65.7\\
& &    20.6 &    9.38 & -- &    75.8 &   118.7 &    41.2 &    61.2\\
&&    41.1 &   10.20 & -- &    77.2 &   121.3 &    39.8 &    58.7\\
&  &    61.7 &    5.65 & -- &    84.7 &   133.1 &    32.2 &    46.9\\
&  &    82.3 &    1.18 & -- &    97.0 &   151.6 &    20.0 &    28.4
  \end{tabular}
 \end{ruledtabular}
 \end{table}
  \begin{table}[ht]
\begin{ruledtabular}
 \begin{tabular}{lcccccccc}
$E_{c.m.}$ & Orientation & $L_0$ & $\tau$ & TKE  & $Z_H$ & $N_H$ & $Z_L$ & $N_L$\\
\colrule
234& 0 & 0 &    9.47 &   216.5 &    83.0 &   129.7 &    34.0 &    50.3\\
&  & 10 &    7.74 &   236.7 &    78.0 &   121.8 &    39.0 &    58.2\\
&  & 20 &    7.17 &   246.1 &    76.8 &   119.3 &    40.2 &    60.7\\
&  & 30 &    6.00 &   223.8 &    80.4 &   126.4 &    36.6 &    53.6\\
&  & 40 &    5.50 &   231.7 &    79.7 &   124.7 &    37.3 &    55.3\\
&  & 50 &    5.47 &   231.0 &    79.5 &   124.4 &    37.5 &    55.6\\
&  & 60 &    5.84 &   226.2 &    79.4 &   123.9 &    37.6 &    56.1\\
&  & 70 &    6.44 &   235.1 &    77.5 &   120.9 &    39.5 &    59.1\\
&  & 80 &    6.87 &   224.8 &    77.7 &   121.3 &    39.3 &    58.7\\
&  & 90 &    7.67 &   223.6 &    77.5 &   120.9 &    39.5 &    59.1\\
&  & 100 &    4.90 &   185.2 &    87.3 &   137.2 &    29.7 &    42.8\\
&  & 110 &    1.37 &   179.2 &    96.3 &   150.4 &    20.7 &    29.6\\
& 15 & 0 &    6.20 &   225.7 &    80.6 &   126.4 &    36.3 &    53.6\\
&  & 10 &    5.84 &   224.9 &    79.2 &   123.9 &    37.8 &    56.1\\
&  & 20 &    5.64 &   236.5 &    79.7 &   124.6 &    37.3 &    55.4\\
&  & 30 &    6.30 &   231.5 &    79.4 &   124.3 &    37.6 &    55.7\\
&  & 40 &    7.47 &   226.7 &    78.4 &   122.2 &    38.6 &    57.8\\
&  & 50 &    8.00 &   225.6 &    78.9 &   123.2 &    38.1 &    56.8\\
&  & 60 &    9.51 &   221.7 &    73.8 &   115.1 &    43.2 &    64.9\\
&  & 70 &    8.20 &   236.1 &    76.2 &   119.1 &    40.8 &    60.9\\
&  & 80 &    9.30 &   203.4 &    74.0 &   115.6 &    43.0 &    64.4\\
&  & 90 &    5.84 &   193.7 &    85.3 &   134.1 &    31.7 &    45.9\\
&  & 100 &    1.03 &   181.8 &    96.7 &   151.0 &    20.3 &    29.0\\
& 30 & 0 &    6.80 &   227.2 &    80.7 &   126.0 &    36.3 &    54.0\\
&  & 10 &    8.07 &   233.6 &    78.2 &   121.8 &    38.8 &    58.2\\
&  & 20 &    9.44 &   231.7 &    79.7 &   124.7 &    37.3 &    55.3\\
&  & 30 &   11.37 &   221.7 &    77.4 &   120.9 &    39.6 &    59.1\\
&  & 40 &   11.84 &   232.7 &    75.5 &   118.2 &    41.5 &    61.8\\
&  & 50 &   11.01 &   229.1 &    74.5 &   116.7 &    42.5 &    63.2\\
&  & 60 &    9.67 &   228.7 &    75.4 &   117.2 &    41.6 &    62.8\\
&  & 70 &    9.74 &   211.0 &    75.1 &   117.7 &    41.9 &    62.3\\
&  & 80 &    8.94 &   204.8 &    79.9 &   124.5 &    37.1 &    55.4\\
&  & 90 &    1.00 &   188.2 &    96.7 &   150.8 &    20.3 &    29.2\\
&  & 100 &    0.13 &   211.3 &    97.1 &   151.9 &    19.9 &    28.1\\
& 45 & 0 &   12.64 &   230.4 &    75.6 &   118.1 &    41.3 &    61.9\\
&  & 10 &   14.34 &   227.4 &    78.5 &   122.5 &    38.5 &    57.5\\
&  & 20 &   13.37 &   227.2 &    79.0 &   123.0 &    38.0 &    57.0\\
&  & 30 &   14.41 &   229.5 &    77.8 &   121.4 &    39.2 &    58.6\\
&  & 40 &   14.61 &   228.2 &    77.1 &   120.3 &    39.9 &    59.7\\
&  & 50 &   11.04 &   217.5 &    75.2 &   117.9 &    41.7 &    62.1\\
&  & 60 &   11.61 &   215.6 &    68.6 &   106.9 &    48.4 &    73.1\\
&  & 70 &   11.47 &   184.5 &    73.4 &   114.6 &    43.6 &    65.4\\
&  & 80 &    2.83 &   171.2 &    94.1 &   147.4 &    22.9 &    32.6\\
&  & 90 &    0.70 &   190.2 &    97.0 &   151.7 &    20.0 &    28.3
  \end{tabular}
 \end{ruledtabular}
 \end{table}
 \begin{table}[ht]
\begin{ruledtabular}
 \begin{tabular}{lcccccccc}
$E_{c.m.}$ & Orientation & $L_0$ & $\tau$ & TKE  & $Z_H$ & $N_H$ & $Z_L$ & $N_L$\\
\colrule
234& 60 & 0 &   10.87 &   239.2 &    78.3 &   122.2 &    38.7 &    57.8\\
&  & 30 &   13.91 &   239.9 &    72.8 &   113.5 &    44.2 &    66.5\\
&  & 40 &   19.04 &   234.1 &    69.6 &   108.7 &    47.4 &    71.3\\
&  & 50 &   12.61 &   228.4 &    70.8 &   110.1 &    46.2 &    69.8\\
&  & 60 &    8.77 &   221.5 &    74.8 &   116.6 &    42.2 &    63.3\\
&  & 70 &    9.37 &   178.4 &    76.1 &   119.3 &    40.9 &    60.7\\
&  & 80 &    5.14 &   180.3 &    87.2 &   136.8 &    29.8 &    43.2\\
&  & 90 &    2.07 &   167.0 &    95.0 &   149.1 &    22.0 &    30.9\\
&  & 100 &    0.40 &   199.0 &    97.2 &   152.0 &    19.8 &    28.0\\
& 75 & 0 &   23.21 &   230.1 &    77.4 &   121.1 &    39.6 &    58.9\\
&  & 40 &   12.11 &   240.0 &    69.9 &   108.5 &    47.1 &    71.5\\
&  & 50 &    8.74 &   225.8 &    74.6 &   116.4 &    42.4 &    63.6\\
&  & 60 &    8.34 &   216.5 &    72.9 &   114.0 &    44.1 &    66.0\\
&  & 70 &   12.57 &   170.7 &    68.3 &   107.0 &    48.7 &    73.0\\
&  & 80 &    5.40 &   185.4 &    86.1 &   135.6 &    30.9 &    44.4\\
&  & 90 &    4.54 &   191.0 &    85.1 &   133.7 &    31.9 &    46.3\\
&  & 100 &    1.33 &   176.7 &    96.0 &   150.4 &    21.0 &    29.6\\
& 90 & 21 &   21.21 &   239.0 &    70.6 &   109.8 &    46.4 &    70.2\\
&  & 32 &   11.97 &   240.4 &    73.9 &   115.4 &    43.1 &    64.6\\
&  & 42 &    8.00 &   222.0 &    77.0 &   120.2 &    40.0 &    59.8\\
&  & 52 &    7.44 &   225.9 &    75.0 &   116.7 &    42.0 &    63.3\\
&  & 60 &    7.60 &   218.4 &    75.3 &   117.4 &    41.7 &    62.6\\
&  & 70 &    7.57 &   193.0 &    78.3 &   122.6 &    38.7 &    57.4\\
&  & 80 &    5.80 &   200.5 &    79.7 &   124.9 &    37.3 &    55.1\\
&  & 90 &    5.40 &   203.5 &    81.7 &   128.3 &    35.3 &    51.7\\
&  & 100 &    4.60 &   192.6 &    83.6 &   131.4 &    33.4 &    48.6\\
&  & 110 &    1.33 &   170.7 &    96.1 &   150.6 &    20.9 &    29.4\\
& 105 & 0 &   23.06 &   229.8 &    77.9 &   121.1 &    39.1 &    58.9\\
&  & 10 &   37.52 &   233.0 &    73.3 &   114.7 &    43.7 &    65.3\\
&  & 20 &    9.27 &   233.2 &    75.9 &   118.8 &    41.1 &    61.2\\
&  & 30 &    8.18 &   227.6 &    77.2 &   120.1 &    39.8 &    59.9\\
&  & 40 &    7.00 &   231.4 &    79.2 &   123.6 &    37.8 &    56.4\\
&  & 50 &    6.45 &   224.7 &    76.9 &   120.2 &    40.1 &    59.8\\
&  & 60 &    5.80 &   228.1 &    77.1 &   120.3 &    39.9 &    59.7\\
&  & 70 &    5.23 &   233.8 &    77.6 &   120.8 &    39.4 &    59.2\\
&  & 80 &    4.77 &   235.8 &    78.2 &   121.8 &    38.8 &    58.2\\
&  & 90 &    4.58 &   232.9 &    78.2 &   122.3 &    38.8 &    57.7\\
&  & 100 &    5.58 &   213.6 &    78.6 &   122.9 &    38.4 &    57.1\\
&  & 110 &    5.03 &   220.6 &    78.7 &   123.2 &    38.3 &    56.8\\
&  & 120 &    3.10 &   157.8 &    94.2 &   147.9 &    22.8 &    32.1\\
&  & 130 &    0.90 &   183.0 &    96.3 &   151.5 &    20.7 &    28.5\\
& 120 & 0 &   10.92 &   239.1 &    78.3 &   122.0 &    38.7 &    58.0\\
&  & 10 &   10.79 &   235.9 &    79.1 &   123.6 &    37.9 &    56.3\\
&  & 20 &    9.08 &   234.4 &    80.2 &   125.1 &    36.8 &    54.8\\
&  & 30 &    7.37 &   219.9 &    79.9 &   125.1 &    37.1 &    54.9\\
&  & 40 &    6.43 &   229.6 &    80.1 &   125.0 &    36.8 &    54.9\\
&  & 50 &    5.87 &   229.7 &    80.4 &   125.8 &    36.6 &    54.2\\
&  & 60 &    5.08 &   229.5 &    79.4 &   124.3 &    37.5 &    55.7\\
&  & 70 &    4.86 &   236.2 &    77.2 &   120.7 &    39.7 &    59.3\\
&  & 80 &    4.82 &   244.3 &    76.5 &   119.6 &    40.5 &    60.4\\
&  & 90 &    4.54 &   244.1 &    76.5 &   119.1 &    40.5 &    60.9\\
&  & 100 &    5.01 &   236.3 &    77.0 &   120.5 &    40.0 &    59.5\\
&  & 110 &    5.56 &   216.1 &    77.7 &   121.8 &    39.3 &    58.2\\
&  & 120 &    5.06 &   223.7 &    81.2 &   127.0 &    35.8 &    53.0\\
&  & 130 &    2.03 &   172.4 &    94.8 &   148.8 &    22.2 &    31.2
 \end{tabular}
 \end{ruledtabular}
 \end{table}
 \begin{table}[ht]
\begin{ruledtabular}
 \begin{tabular}{lcccccccc}
$E_{c.m.}$ & Orientation & $L_0$ & $\tau$ & TKE  & $Z_H$ & $N_H$ & $Z_L$ & $N_L$\\
\colrule
234& 135 & 0 &   12.50 &   230.6 &    75.6 &   118.1 &    41.3 &    61.9\\
&  & 10 &    8.50 &   222.4 &    80.3 &   125.3 &    36.7 &    54.7\\
&  & 20 &    7.58 &   227.7 &    79.6 &   124.4 &    37.4 &    55.6\\
&  & 30 &    6.95 &   229.6 &    79.9 &   124.8 &    37.1 &    55.2\\
&  & 40 &    5.23 &   234.9 &    80.1 &   125.3 &    36.9 &    54.7\\
&  & 50 &    4.76 &   232.5 &    79.4 &   124.2 &    37.6 &    55.8\\
&  & 60 &    4.80 &   226.3 &    79.4 &   123.9 &    37.6 &    56.1\\
&  & 70 &    5.12 &   236.3 &    76.9 &   119.6 &    40.1 &    60.4\\
&  & 80 &    5.64 &   235.1 &    77.4 &   120.1 &    39.6 &    59.9\\
&  & 90 &    5.52 &   239.1 &    78.4 &   122.4 &    38.6 &    57.6\\
&  & 100 &    5.89 &   236.3 &    78.1 &   121.4 &    38.9 &    58.6\\
&  & 110 &    5.65 &   226.4 &    79.1 &   123.0 &    37.9 &    56.9\\
&  & 120 &    5.30 &   202.9 &    85.3 &   133.8 &    31.7 &    46.2\\
&  & 130 &    1.80 &   176.2 &    95.1 &   149.2 &    21.9 &    30.8\\
& 150 & 0 &    6.72 &   227.7 &    80.7 &   126.0 &    36.3 &    53.9\\
&  & 10 &    5.98 &   228.7 &    79.9 &   125.0 &    37.1 &    55.0\\
&  & 20 &    5.16 &   236.9 &    80.2 &   125.3 &    36.8 &    54.7\\
&  & 30 &    5.25 &   232.0 &    80.2 &   125.1 &    36.8 &    54.9\\
&  & 40 &    6.03 &   231.3 &    78.5 &   122.9 &    38.5 &    57.1\\
&  & 50 &    5.94 &   233.8 &    79.5 &   123.8 &    37.5 &    56.2\\
&  & 60 &    6.39 &   220.6 &    80.2 &   125.2 &    36.8 &    54.8\\
&  & 70 &    6.62 &   227.7 &    78.9 &   122.9 &    38.1 &    57.1\\
&  & 80 &    6.06 &   240.1 &    78.3 &   121.7 &    38.8 &    58.3\\
&  & 90 &    5.69 &   242.4 &    77.5 &   120.5 &    39.5 &    59.5\\
&  & 100 &    5.82 &   234.8 &    77.8 &   121.3 &    39.2 &    58.7\\
&  & 110 &    5.90 &   208.3 &    82.5 &   128.9 &    34.4 &    51.0\\
&  & 120 &    2.80 &   168.2 &    94.3 &   148.1 &    22.7 &    31.9\\
&  & 130 &    0.90 &   186.0 &    96.8 &   151.4 &    20.2 &    28.6\\
& 165 & 0 &    6.27 &   225.6 &    80.6 &   126.4 &    36.4 &    53.6\\
&  & 10 &    6.86 &   219.5 &    81.0 &   126.7 &    36.0 &    53.3\\
&  & 20 &    7.06 &   225.1 &    79.4 &   123.9 &    37.6 &    56.1\\
&  & 30 &    7.59 &   232.4 &    79.0 &   123.4 &    38.0 &    56.6\\
&  & 40 &    7.36 &   238.8 &    77.4 &   119.8 &    39.6 &    60.2\\
&  & 50 &    6.86 &   242.4 &    77.6 &   120.5 &    39.4 &    59.5\\
&  & 60 &    5.90 &   230.8 &    79.1 &   123.7 &    37.9 &    56.3\\
&  & 70 &    5.84 &   231.5 &    78.5 &   122.1 &    38.5 &    57.9\\
&  & 80 &    5.87 &   234.0 &    76.8 &   119.7 &    40.2 &    60.3\\
&  & 90 &    5.70 &   225.1 &    79.6 &   124.5 &    37.4 &    55.5\\
&  & 100 &    4.59 &   201.7 &    80.1 &   125.6 &    36.9 &    54.4\\
&  & 110 &    4.43 &   180.4 &    89.9 &   141.0 &    27.1 &    39.0\\
&  & 120 &    1.17 &   183.1 &    96.2 &   150.5 &    20.8 &    29.5\\
& 180 & 0 &    9.47 &   216.5 &    83.0 &   129.7 &    34.0 &    50.3\\
&  & 10 &    7.74 &   236.7 &    78.0 &   121.8 &    39.0 &    58.2\\
&  & 20 &    7.17 &   246.1 &    76.8 &   119.3 &    40.2 &    60.7\\
&  & 30 &    6.00 &   223.8 &    80.4 &   126.4 &    36.6 &    53.6\\
&  & 40 &    5.50 &   231.7 &    79.7 &   124.7 &    37.3 &    55.3\\
&  & 50 &    5.47 &   231.0 &    79.5 &   124.4 &    37.5 &    55.6\\
&  & 60 &    5.84 &   226.2 &    79.4 &   123.9 &    37.6 &    56.1\\
&  & 70 &    6.44 &   235.1 &    77.5 &   120.9 &    39.5 &    59.1\\
&  & 80 &    6.87 &   224.8 &    77.7 &   121.3 &    39.3 &    58.7\\
&  & 90 &    7.67 &   223.6 &    77.5 &   120.9 &    39.5 &    59.1\\
&  & 100 &    4.90 &   185.2 &    87.3 &   137.2 &    29.7 &    42.8\\
&  & 110 &    1.37 &   179.2 &    96.3 &   150.4 &    20.7 &    29.6
  \end{tabular}
 \end{ruledtabular}
 \end{table}

%------------------------------------------------------------------------------

\end{document}